\documentclass[11pt]{article}
\usepackage{amssymb,amsmath,amsfonts}
\usepackage{graphicx}
\usepackage{graphics}
\usepackage{eepic,epsfig}

\@addtoreset{equation}{section}

\makeatother

\begin{document}

\title{Fermionic condensate and Casimir densities in the presence \\
of compact dimensions with applications to nanotubes}
\author{E. Elizalde$^{1}$,\thinspace\ S.~D. Odintsov$^{1,2}$\thanks{%
Also at Tomsk State Pedagogical University, Tomsk}, \thinspace\ A.~A.
Saharian$^{3}$ \\
\\
\textit{$^{1}$Instituto de Ciencias del Espacio (CSIC) }\\
\textit{and Institut d'Estudis Espacials de Catalunya (IEEC/CSIC) }\\
\textit{Campus UAB, Facultat de Ci\`{e}ncies, Torre C5-Parell-2a planta,}\\
\textit{08193 Bellaterra (Barcelona) Spain}\vspace{0.3cm}\\
\textit{$^2$Instituci\'{o} Catalana de Recerca i Estudis Avan\c{c}ats (ICREA)%
}\vspace{0.3cm}\\
\textit{$^3$Department of Physics, Yerevan State University,}\\
\textit{1 Alex Manoogian Street, 0025 Yerevan, Armenia}}
\maketitle

\begin{abstract}
We investigate the fermionic condensate and the vacuum expectation value of
the energy-momentum tensor for a massive fermionic field in the geometry of
two parallel plate on the background of Minkowski spacetime with an
arbitrary number of toroidally compactified spatial dimensions, in the
presence of a constant gauge field. Bag boundary conditions are imposed on
the plates and periodicity conditions with arbitrary phases are considered
along the compact dimensions. The nontrivial topology of the background
spacetime leads to an Aharonov-Bohm effect for the vacuum expectation values
induced by the gauge field. The fermionic condensate and the expectation
value of the energy-momentum tensor are periodic functions of the magnetic
flux with period equal to the flux quantum. The boundary induced parts in
the fermionic condensate and the vacuum energy density are negative, with
independence of the phases in the periodicity conditions and of the value of
the gauge potential. Interaction forces between the plates are thus always
attractive.

However, in physical situations where the quantum field is confined to the
region between the plates, the pure topological part contributes as well,
and then the resulting force can be either attractive or repulsive,
depending on the specific phases encoded in the periodicity conditions along
the compact dimensions, and on the gauge potential, too. Applications of the
general formulas to cylindrical carbon nanotubes are considered, within the
framework of a Dirac-like theory for the electronic states in graphene. In
the absence of a magnetic flux, the energy density for semiconducting
nanotubes is always negative. For metallic nanotubes the energy density is
positive for long tubes and negative for short ones. The resulting Casimir
forces acting on the edges of the nanotube are attractive for short tubes
with independence of the tube chirality. The sign of the force for long
nanotubes can be controlled by tuning the magnetic flux. This opens the way
to the design of efficient actuators driven by the Casimir force at the
nanoscale.
\end{abstract}

\bigskip

\section{Introduction}

\label{sec:Introd}

In a good number of problems one needs to consider a physical model on the
background of some manifold with compactified spatial dimensions. Many
high-energy theories of fundamental physics are formulated in a higher
dimensional spacetime and it is commonly assumed that the extra dimensions
are compactified. In particular, additional compact dimensions have been
extensively used in supergravity and superstring theories. From the
inflationary point of view, universes with compact dimensions, under certain
conditions, should be considered a rule rather than an exception \cite%
{Lind04}. Models of a compact universe with non-trivial topology may play an
important role by providing proper initial conditions for inflation. There
has been a large activity to search for signatures of non-trivial topology
by identifying ghost images of galaxies, clusters or quasars. Recent
progress in observations of the cosmic microwave background provides an
alternative way to observe the topology of the universe. An interesting
application of the field theoretical models with compact dimensions recently
appeared in nanophysics \cite{Sait98}. The long-wavelength description of
the electronic states in graphene can be formulated in terms of the
Dirac-like theory in 3-dimensional spacetime with the Fermi velocity playing
the role of speed of light (see, e.g., Refs.~\cite{Seme84,Vinc84}).
Single-walled carbon nanotubes are generated by rolling up a graphene sheet
to form a cylinder and the background spacetime for the corresponding
Dirac-like theory has topology $R^{2}\times S^{1}$. \ For another class of
graphene-made structures, called toroidal carbon nanotubes, one has as
background topology $R^{1}\times (S^{1})^{2}$.

The boundary conditions imposed on fields along compact dimensions give rise
to a modification of the spectrum of the vacuum fluctuations and, as a
result, to Casimir-type contributions in the vacuum expectation values of
physical observables (for the topological Casimir effect and its role in
cosmology see \cite{Most97} and references therein). In models of
Kaluza-Klein type, the Casimir effect has been used as a stabilization
mechanism for moduli fields and as a source for dynamical compactification
of the extra dimensions, in particular, for quantum Kaluza-Klein gravity
(see Ref. \cite{Buch89}). The Casimir energy can also serve as a model for
dark energy needed for the explanation of the present accelerated expansion
of the universe (see \cite{Eliz01} and references therein). In addition,
recent measurements of the Casimir forces between macroscopic bodies provide
a sensitive test for constraining the parameters of long-range interactions,
as predicted by modern unification theories of fundamental interactions \cite%
{Most87}. The influence of extra compactified dimensions on the Casimir
effect in the classical configuration of two parallel plates has been
recently discussed in \cite{Chen06}, for the case of a scalar field, and in
\cite{Popp04}, for the electromagnetic field with perfectly conducting
boundary conditions.

More recently, interest has been focussed on the topic of the Casimir effect
in braneworld models with large extra dimensions. This type of models (for a
review see \cite{Brane}) naturally appear in the string/M theory context and
they provide a novel set up for discussing phenomenological and cosmological
issues related with extra dimensions. In braneworld models the investigation
of quantum effects is of considerable phenomenological value, both in
particle physics and in cosmology. The braneworld corresponds to a manifold
with boundaries and the bulk fields will give Casimir-type contributions to
the vacuum energy and, as a result, to the vacuum forces acting on the
branes. The Casimir forces provide a natural mechanism for stabilizing the
radion field in the Randall-Sundrum model, as required for a complete
solution of the hierarchy problem. In addition, the Casimir energy gives a
contribution to both the brane and the bulk cosmological constants. Hence,
it has to be taken into account in any self-consistent formulation of the
braneworld dynamics. The Casimir energy and corresponding Casimir forces
within the framework of the Randall-Sundrum braneworld \cite{Rand99} have
been evaluated in Refs.~\cite{Gold00} by using both dimensional and zeta
function regularization methods. Local Casimir densities were considered in
Ref.~\cite{Knap04}. The Casimir effect in higher dimensional generalizations
of the Randall-Sundrum model with compact internal spaces has been
investigated in \cite{Flac03}.

In the present paper we will study the fermionic condensate, the Casimir
energy density and the vacuum stresses for a massive fermion field in the
geometry of two parallel plates on a spacetime with an arbitrary number of
toroidally compactified spatial dimensions. We will impose generalized
periodicity conditions along the compact dimensions with arbitrary phases
and MIT bag boundary conditions on the plates. The presence of a constant
gauge field will be assumed as well. Though the corresponding field strength
vanishes, the nontrivial topology of the background spacetime leads to
Aharonov-Bohm-like effects on the vacuum expectation values. The total
Casimir energy in the geometry under consideration has been discussed in
Ref.~\cite{Bell09}, in the absence of a gauge field. The investigation of
local physical characteristics in the Casimir effect, such as the
expectation value of the energy-momentum tensor and the fermionic
condensate, is of considerable interest. Indeed, local quantities contain
more information on the vacuum fluctuations than global ones. In addition to
describing the physical structure of the quantum field at a given point, the
energy-momentum tensor acts as the source in Einstein's equations and,
therefore, it plays an important role in modeling a self-consistent dynamics
which involves the gravitational field. The fermionic condensate plays an
important role in the models of dynamical chiral symmetry breaking (see
review~\cite{Inag97} for chiral symmetry breaking in the Nambu-Jona-Lasino
and Gross-Neveu models on the background of a curved spacetime with
non-trivial topology and \cite{Flac10} for very recent developments).

The fermion Casimir energy for two parallel plates in 4-dimensional
Minkowski spacetime with trivial topology has been considered in \cite%
{John75}, for a massless field, and in \cite{Mama80} in the massive case.
For arbitrary number of dimensions, the corresponding results are
generalized in Refs.~\cite{Paol99,Eliz02} for the massless and massive
cases, respectively. The fermionic condensate for a massless field has been
considered in Refs.~\cite{Lutk84}. The Casimir problem for fermions coupled
to a static background field in one spatial dimension is investigated in
\cite{Sund04}. The interaction energy density and the corresponding force
are computed in the limit that the background becomes concentrated at two
points. The fermionic Casimir effect for parallel plates with imperfect bag
boundary conditions modelled by $\delta $-like potentials is studied in \cite%
{Fosc08}. The topological Casimir effect and the vacuum expectation value of
the fermionic current for a massive fermionic field in a spacetime with an
arbitrary number of toroidally compactified spatial dimensions have been
considered in \cite{Bell09b,Bell10}.

The paper is organized as follows. In the next section, we specify the
eigenfunctions and the eigenmodes for the Dirac equation in the region
between the plates assuming bag boundary conditions on them. The fermionic
condensate in this region is considered in Sect.~\ref{sec:FC}. By using an
Abel-Plana-type summation formula, we express the condensate as the sum of a
pure topological, single plate contribution and interference parts. Various
limiting cases are then considered. The vacuum expectation value of the
energy-momentum tensor is investigated in Sect.~\ref{sec:EMT}. A
generalization for a conformally-flat background spacetime is given. In
Sect.~\ref{sec:Nano} we study applications of the general formulas to the
Casimir effect for electrons in a carbon nanotube, within the framework of a
3-dimensional Dirac-like model. The main results of the paper are summarized
in Sect.~\ref{sec:Conc}.

\section{Fermionic eigenfunctions}

\label{sec:Modes}

Consider a spinor field, $\psi $, propagating on a $(D+1)$-dimensional flat
spacetime with spatial topology $R^{p+1}\times (S^{1})^{q}$, $p+q+1=D$. We
will denote by $\mathbf{z}_{p+1}=(z_{1},\ldots ,z_{p+1}\equiv z)$ and $%
\mathbf{z}_{q}=(z_{p+2},\ldots ,z_{D})$ the Cartesian coordinates along the
uncompactified and the compactified dimensions, respectively. For these
coordinates we have $-\infty <z_{l}<\infty $, $l=1,\ldots ,p+1$, and $%
0\leqslant z_{l}\leqslant L_{l}$ for $l=p+2,\ldots ,D$, with $L_{l}$ being
the length of the $l$-th compact dimension. We assume that along the
compactified dimensions the field obeys quasiperiodic boundary conditions%
\begin{equation}
\psi (t,\mathbf{z}_{p+1},\mathbf{z}_{q}+L_{l}\mathbf{e}_{l})=e^{2\pi i\alpha
_{l}}\psi (t,\mathbf{z}_{p+1},\mathbf{z}_{q}),  \label{BC}
\end{equation}%
with constant phases $|\alpha _{l}|\leqslant 1/2$ and $\mathbf{e}_{l}$\ is
the unit vector along the direction of the coordinate $z_{l}$, $l=p+2,\ldots
,D$. Condition (\ref{BC}) includes the periodicity conditions for both
untwisted and twisted fermionic fields as special cases with $\alpha _{l}=0$
and $\alpha _{l}=1/2$, respectively. The special cases $\alpha _{l}=0,\pm
1/3 $ are realized in nanotubes.

In this paper we are interested in the fermionic condensate and in the
vacuum expectation value (VEV) of the energy-momentum tensor induced by two
parallel plates located at $z=0$ and $z=a$. On the boundaries the field
obeys MIT bag boundary conditions
\begin{equation}
\left( 1+i\gamma ^{\mu }n_{\mu }\right) \psi =0\ ,\quad z=0,a,
\label{BagCond}
\end{equation}%
with $\gamma ^{\mu }$ being the Dirac matrices and $n_{\mu }$ the outward
oriented (with respect to the region under consideration) normal to the
boundary. Note that from the conditions (\ref{BagCond}) it follows that, on
the boundaries, $\bar{\psi}\psi =0$ and $n_{\mu }\bar{\psi}\gamma ^{\mu
}\psi =0$, where $\bar{\psi}=\psi ^{\dagger }\gamma ^{0}$ is the Dirac
adjoint and the dagger denotes Hermitian conjugation. In the discussion
below the calculations will be done for the region between the plates, $0<z<a
$, where we have $n_{\mu }=-\delta _{\mu }^{p+1}$ at $z=0$ and $n_{\mu
}=\delta _{\mu }^{p+1}$ at $z=a$. The expressions for the VEVs in the
regions $z<0$ and $z>a $ are obtained as limiting cases.

Dynamics of the massive spinor field is governed by the Dirac equation
\begin{equation}
i\gamma ^{\mu }\partial _{\mu }\psi -m\psi =0\ .  \label{Direq}
\end{equation}%
In the $(D+1)$-dimensional spacetime, the Dirac matrices are $N_{D}\times
N_{D}$ matrices with $N_{D}=2^{[(D+1)/2]}$, where the square brackets mean
integer part of the enclosed expression. We will take these matrices in the
Dirac representation:
\begin{equation}
\gamma ^{0}=\left(
\begin{array}{cc}
1 & 0 \\
0 & -1%
\end{array}%
\right) ,\;\gamma ^{\mu }=\left(
\begin{array}{cc}
0 & \sigma _{\mu } \\
-\sigma _{\mu }^{+} & 0%
\end{array}%
\right) ,\;\mu =1,2,\ldots ,D.  \label{DiracMat}
\end{equation}%
From the anticommutation relations for the Dirac matrices one has $\sigma
_{\mu }\sigma _{\nu }^{+}+\sigma _{\nu }\sigma _{\mu }^{+}=2\delta _{\mu \nu
}$. In the case $D=2$ we have $N_{D}=2$ and the Dirac matrices are  $\gamma
^{\mu }=(\sigma _{\text{P}3},i\sigma _{\text{P}1},i\sigma _{\text{P}2})$,
with $\sigma _{\text{P}\mu }$ being the $2\times 2$ Pauli matrices.

The boundary conditions (\ref{BC}) and (\ref{BagCond}) lead to the
modification of the spectrum for vacuum fluctuations of the fermionic field
and, as a result, to the topological and boundary induced Casimir effects on
the VEVs of physical observables. For the evaluation of the VEVs we need the
complete set of positive- and negative-energy solutions to the Dirac
equation satisfying the boundary conditions (\ref{BagCond}). The dependence
on the coordinates parallel to the plates, $\mathbf{z}_{\parallel }=$ $%
(z_{1},\ldots ,z_{p},z_{p+2},\ldots ,z_{D})$, can be presented in the
standard exponential form $\exp (i\mathbf{k}_{\parallel }\cdot \mathbf{z}%
_{\parallel })$, with $\mathbf{k}_{\parallel }=(\mathbf{k}_{p},\mathbf{k}%
_{q})$ and $\mathbf{k}_{p}=(k_{1},\ldots ,k_{p})$, $\mathbf{k}%
_{q}=(k_{p+2},\ldots ,k_{D})$. The eigenvalues for the components of the
wave vector along the compactified dimensions are determined from the
periodicity conditions (\ref{BC}):%
\begin{equation}
\mathbf{k}_{q}=(2\pi (n_{p+2}+\alpha _{p+2})/L_{p+2},\ldots ,2\pi
(n_{D}+\alpha _{D})/L_{D}),  \label{kDn}
\end{equation}%
with $n_{p+2},\ldots ,n_{D}=0,\pm 1,\pm 2,\ldots $. For the components along
the uncompactified dimensions one has $-\infty <k_{l}<\infty $, $l=1,\ldots
,p$. The corresponding positive- and negative-energy eigenspinors have the
form
\begin{eqnarray}
\psi _{\beta }^{(+)} &=&A_{\beta }e^{-i\omega t}\left(
\begin{array}{c}
\varphi \\
-i\boldsymbol{\sigma }^{+}\cdot \boldsymbol{\nabla }\varphi /\left( \omega
+m\right)%
\end{array}%
\right) ,  \notag \\
\psi _{\beta }^{(-)} &=&A_{\beta }e^{i\omega t}\left(
\begin{array}{c}
i\boldsymbol{\sigma }\cdot \boldsymbol{\nabla }\chi /\left( \omega +m\right)
\\
\chi%
\end{array}%
\right) ,  \label{Eigfunc}
\end{eqnarray}%
where $\boldsymbol{\sigma }=(\sigma _{1},\ldots ,\sigma _{D})$, $\omega =%
\sqrt{\mathbf{k}_{p}^{2}+k_{p+1}^{2}+\mathbf{k}_{q}^{2}+m^{2}}$ and $\beta $
is the collective index for the set of quantum numbers specifying the
solutions (see below). The spinors in (\ref{Eigfunc}) are given by the
expressions%
\begin{eqnarray}
\varphi &=&e^{i\mathbf{k}_{\parallel }\cdot \mathbf{z}_{\parallel }}\left(
\varphi _{+}e^{ik_{p+1}z}+\varphi _{-}e^{-ik_{p+1}z}\right) ,  \notag \\
\chi &=&e^{-i\mathbf{k}_{\parallel }\cdot \mathbf{z}_{\parallel }}\left(
\chi _{+}e^{ik_{p+1}z}+\chi _{-}e^{-ik_{p+1}z}\right) ,  \label{phixi}
\end{eqnarray}

From the boundary condition (\ref{BagCond}) on the plate at $z=0$ we find
the following relations between the spinors in (\ref{phixi})%
\begin{eqnarray}
\varphi _{+} &=&-\frac{m(\omega +m)+k_{p+1}^{2}-k_{p+1}\sigma _{p+1}%
\boldsymbol{\sigma }_{\parallel }^{+}\cdot \mathbf{k}_{\parallel }}{%
(m-ik_{p+1})\left( \omega +m\right) }\varphi _{-},  \notag \\
\chi _{-} &=&-\frac{m(\omega +m)+k_{p+1}^{2}-k_{p+1}\sigma _{p+1}^{+}%
\boldsymbol{\sigma }_{\parallel }\cdot \mathbf{k}_{\parallel }}{%
(m+ik_{p+1})(\omega +m)}\chi _{+},  \label{phixiRel}
\end{eqnarray}%
where $\boldsymbol{\sigma }_{\parallel }=(\sigma _{1},\ldots ,\sigma
_{p},\sigma _{p+2},\ldots ,\sigma _{D})$. We will assume that they are
normalized in accordance with $\varphi _{-}^{+}\varphi _{-}=\chi
_{+}^{+}\chi _{+}=1$. As a set of independent spinors we will take $\varphi
_{-}=w^{(\sigma )}$ and $\chi _{+}=w^{(\sigma )\prime }$, where $w^{(\sigma
)}$, $\sigma =1,\ldots ,N_{D}/2$, are one-column matrices having $N_{D}/2$
rows with the elements $w_{l}^{(\sigma )}=\delta _{l\sigma }$, and $%
w^{(\sigma )\prime }=iw^{(\sigma )}$. Now the set of quantum numbers
specifying the eigenfunctions (\ref{Eigfunc}) is $\beta =(\mathbf{k},\sigma
) $. From the boundary condition at $z^{p+1}=a$ it follows that the
eigenvalues of $k_{p+1}$ are roots of the transcendental equation
\begin{equation}
ma\sin (k_{p+1}a)/(k_{p+1}a)+\cos (k_{p+1}a)=0.  \label{kpvalues}
\end{equation}%
All these roots are real. We will denote the positive solutions of Eq.~(\ref%
{kpvalues}) by $\lambda _{n}=k_{p+1}a$, $n=1,2,\ldots $. For a massless
field one has $\lambda _{n}=\pi (n-1/2)$. Note that Eq.~(\ref{kpvalues})
does not contain the parameters of the compact subspace and is the same as
in the corresponding problem on the topologically trivial Minkowski
spacetime (see \cite{Most97}). Note that this will not be the case in a more
general class of compact subspaces.

The normalization coefficient $A_{\beta }$ in (\ref{Eigfunc}) is determined
from the orthonormalization condition%
\begin{equation}
\int d\mathbf{z}_{\parallel }\int_{0}^{a}dz^{p+1}\,\psi _{\beta }^{(\pm
)+}\psi _{\beta ^{\prime }}^{(\pm )}=\delta _{\beta \beta ^{\prime }}.
\label{normaliz}
\end{equation}%
Here, the symbol $\delta _{\beta \beta ^{\prime }}$ is understood as the
Dirac delta function for continuous indices and the Kronecker delta for
discrete ones. The substitution of the eigenfunctions (\ref{Eigfunc}) into
this condition leads to the result%
\begin{equation}
A_{\beta }^{2}=\frac{\omega +m}{4(2\pi )^{p}\omega aV_{q}}\left[ 1-\frac{%
\sin (2k_{p+1}a)}{2k_{p+1}a}\right] ^{-1},  \label{normcoef}
\end{equation}%
where $V_{q}=L_{p+2}\cdots L_{D}$ is the volume of the compact subspace.

We can generalize the eigenfunctions given above to the situation when an
external electromagnetic field with vector potential $A_{\mu }=\mathrm{const}
$ is present. In spite of the fact that the corresponding magnetic field
strength vanishes, the non-trivial topology of the background spacetime
leads to the appearance of an Aharonov-Bohm-like effect for the physical
observables. In particular, the corresponding VEVs depend on $A_{\mu }$. Now
the Dirac equation has the form $i\gamma ^{\mu }(\partial _{\mu }+ieA_{\mu
})\psi -m\psi =0$ and, by making use the gauge transformation $A_{\mu
}=A_{\mu }^{\prime }+\partial _{\mu }\Lambda (x)$, $\psi (x)=\psi ^{\prime
}(x)e^{-ie\Lambda (x)}$, with the function $\Lambda (x)=A_{\mu }x^{\mu }$,
we see that the new function $\psi ^{\prime }(x)$ satisfies the Dirac
equation with $A_{\mu }^{\prime }=0$ and the quasiperiodicity conditions
similar to (\ref{BC}) with the replacement%
\begin{equation}
\alpha _{l}\rightarrow \tilde{\alpha}_{l}=\alpha _{l}+eA_{l}L_{l}/(2\pi ).
\label{PhaseRepl}
\end{equation}%
The eigenvalues for the wave vector components along compact dimensions are
defined by $k_{l}=2\pi (n_{l}+\tilde{\alpha}_{l})/L_{l}$ and the
corresponding eigenspinors are obtained from those given above with the
replacement (\ref{PhaseRepl}).

\section{Fermionic condensate}

\label{sec:FC}

The fermionic condensate is among the most important quantities that
characterize the properties of the quantum vacuum. Although the
corresponding operator is local, due to the global nature of the vacuum,
this quantity carries important information about the global properties of
the background spacetime. Having the complete set of eigenspinors, we can
evaluate the fermionic condensate by using the mode-sum%
\begin{equation}
\langle \bar{\psi}\psi \rangle =\sum_{\beta }\bar{\psi}_{\beta
}^{(-)}(x)\psi _{\beta }^{(-)}(x),  \label{FC1}
\end{equation}%
where $\langle \cdots \rangle $ stands for VEV. By taking into account the
expression (\ref{Eigfunc}) for the negative-energy eigenspinors, the
fermionic condensate in the region between the plates can be expressed as%
\begin{eqnarray}
\langle \bar{\psi}\psi \rangle &=&-\frac{N_{D}}{aV_{q}}\sum_{\mathbf{n}%
_{q}\in \mathbf{Z}^{q}}\int \frac{d\mathbf{k}_{p}}{(2\pi )^{p}}%
\sum_{n=1}^{\infty }\frac{\sin (\lambda _{n}z/a)}{\omega a}  \notag \\
&&\times \frac{ma\sin (\lambda _{n}z/a)+\lambda _{n}\cos (\lambda _{n}z/a)}{%
1-\sin (2\lambda _{n})/(2\lambda _{n})},  \label{FC2}
\end{eqnarray}%
where $\mathbf{n}_{q}=(n_{p+2},\ldots ,n_{D})$ and%
\begin{equation}
\omega =\sqrt{\lambda _{n}^{2}/a^{2}+\mathbf{k}_{p}^{2}+\mathbf{k}%
_{q}^{2}+m^{2}}.  \label{omeg}
\end{equation}
Of course, the expression on the rhs of Eq.~(\ref{FC2}) is divergent. We
will assume that some cutoff function is present, without writing it
explicitly.

For the further evaluation of the fermionic condensate we apply to the sum
over $n$ in Eq.~(\ref{FC2}) the Abel-Plana-type summation formula%
\begin{equation}
\sum_{n=1}^{\infty }\frac{\pi f(\lambda _{n})}{1-\sin (2\lambda
_{n})/(2\lambda _{n})}=-\frac{\pi maf(0)}{2(ma+1)}+\int_{0}^{\infty
}dx\,f(x)-i\int_{0}^{\infty }dx\frac{f(ix)-f(-ix)}{\frac{x+ma}{x-ma}e^{2x}+1}%
,  \label{Abel-Plan}
\end{equation}%
with the function%
\begin{equation}
f(x)=\frac{\sin (xz/a)}{\sqrt{x^{2}+\mathbf{k}_{p}^{2}a^{2}+m_{\mathbf{n}%
_{q}}^{2}a^{2}}}[ma\sin (xz/a)+x\cos (xz/a)].  \label{fz}
\end{equation}%
Formula (\ref{Abel-Plan}) is a special case of the summation formula derived
in Ref.~\cite{Rome02} on the basis of the generalized Abel-Plana formula
(see also Ref.~\cite{Saha08Rev}). In Eq.~(\ref{fz}) and the discussion below
we use the notation%
\begin{equation}
m_{\mathbf{n}_{q}}^{2}=k_{\mathbf{n}_{q}}^{2}+m^{2},\;k_{\mathbf{n}%
_{q}}^{2}=\sum_{l=p+2}^{D}[2\pi (n_{l}+\alpha _{l})/L_{l}]^{2}.  \label{mnq}
\end{equation}%
After the application of formula (\ref{Abel-Plan}), the fermionic condensate
is split into%
\begin{eqnarray}
\langle \bar{\psi}\psi \rangle  &=&\langle \bar{\psi}\psi \rangle
^{(0)}+\langle \bar{\psi}\psi \rangle ^{(1)}-\frac{2N_{D}}{\pi V_{q}}\sum_{%
\mathbf{n}_{q}\in \mathbf{Z}^{q}}\int \frac{d\mathbf{k}_{p}}{(2\pi )^{p}}%
\int_{\sqrt{\mathbf{k}_{p}^{2}+m_{\mathbf{n}_{q}}^{2}}}^{\infty }dx  \notag
\\
&&\times \frac{\sinh (xz)}{\sqrt{x^{2}-\mathbf{k}_{p}^{2}-m_{\mathbf{n}%
_{q}}^{2}}}\frac{m\sinh (xz)+x\cosh (xz)}{\frac{x+m}{x-m}e^{2ax}+1},
\label{FC3}
\end{eqnarray}%
where%
\begin{equation}
\langle \bar{\psi}\psi \rangle ^{(0)}=-\frac{N_{D}m}{2V_{q}}\sum_{\mathbf{n}%
_{q}\in \mathbf{Z}^{q}}\int \frac{d\mathbf{k}_{p+1}}{(2\pi )^{p+1}}\frac{1}{%
\sqrt{\mathbf{k}_{p+1}^{2}+m_{\mathbf{n}_{q}}^{2}}},  \label{FC(0)}
\end{equation}%
is the fermionic condensate in the topology $R^{p+1}\times (S^{1})^{q}$ when
the boundaries are absent. The term%
\begin{equation}
\langle \bar{\psi}\psi \rangle ^{(1)}=-\frac{N_{D}}{2\pi V_{q}}\sum_{\mathbf{%
n}_{q}\in \mathbf{Z}^{q}}\int \frac{d\mathbf{k}_{p}}{(2\pi )^{p}}%
\int_{0}^{\infty }dx\,\frac{x\sin (2xz)-m\cos (2xz)}{\sqrt{x^{2}+\mathbf{k}%
_{p}^{2}+m_{\mathbf{n}_{q}}^{2}}},  \label{FC(1)}
\end{equation}%
is the part induced by the plate at $z=0$ when the second plate is absent.
The last term on the right of formula (\ref{FC3}) comes from the last term
in Eq.~(\ref{Abel-Plan}) and it is induced by the presence of the second
plate. Note that this term vanishes at $z=0$.

For points away from the boundaries, the boundary induced part is finite and
the cutoff function in the corresponding expressions can be safely removed.
Renormalization is needed for the purely topological part only. The latter
has been investigated in Ref.~\cite{Bell09b} (for the topological fermionic
Casimir effect in de Sitter spacetime with toroidally compactified spatial
dimensions see Ref.~\cite{Saha08}). By making use of the zeta function
technique, the corresponding renormalized expression is presented in the form%
\begin{eqnarray}
\langle \bar{\psi}\psi \rangle ^{(0)} &=&-\frac{N_{D}m}{(2\pi )^{(D+1)/2}}%
\sideset{}{'}{\sum}_{\mathbf{n}_{q}\in \mathbf{Z}^{q}}\cos (2\pi \mathbf{n}%
_{q}\cdot \boldsymbol{\alpha }_{q})  \notag \\
&&\times \frac{f_{(D-1)/2}(m\sqrt{L_{p+2}^{2}n_{p+2}^{2}+\cdots
+L_{D}^{2}n_{D}^{2}})}{(L_{p+2}^{2}n_{p+2}^{2}+\cdots
+L_{D}^{2}n_{D}^{2})^{(D-1)/2}},  \label{FC0b}
\end{eqnarray}%
with $\boldsymbol{\alpha }_{q}=(\alpha _{p+2},\ldots ,\alpha _{D})$, and%
\begin{equation}
f_{\nu }(z)=z^{\nu }K_{\nu }(z).  \label{fnu}
\end{equation}%
The prime on the summation sign in (\ref{FC0b}) means that the term $\mathbf{%
n}_{q}=0$ is excluded from the sum. An alternative expression for $\langle
\bar{\psi}\psi \rangle ^{(0)}$ is obtained in Ref.~\cite{Bell09b} using the
Abel-Plana summation formula. In the discussion below we will be
concentrated on the boundary induced parts.

For further transformation of the single plate part (\ref{FC(1)}), we write
the function in the integrand as%
\begin{equation}
x\sin (2xz)-m\cos (2xz)=-\frac{1}{2}\left[ \left( m+ix\right)
e^{2ixz}+\left( m-ix\right) e^{-2ixz}\right] .  \label{rel1}
\end{equation}%
In the integral over $x$ in (\ref{FC(1)}) we rotate the integration contour
by an angle $\pi /2$, for the term with the exponent $e^{2ixz}$, and by $%
-\pi /2$, for the term with the exponent $e^{-2ixz}$. As a result, we get%
\begin{equation}
\langle \bar{\psi}\psi \rangle ^{(1)}=\frac{N_{D}}{2\pi V_{q}}\sum_{\mathbf{n%
}_{q}\in \mathbf{Z}^{q}}\int \frac{d\mathbf{k}_{p}}{(2\pi )^{p}}\int_{\sqrt{%
\mathbf{k}_{p}^{2}+m_{\mathbf{n}_{q}}^{2}}}^{\infty }dx\frac{\left(
m-x\right) e^{-2xz}}{\sqrt{x^{2}-\mathbf{k}_{p}^{2}-m_{\mathbf{n}_{q}}^{2}}}.
\label{FC(1)1}
\end{equation}
It follows from this expression that $\langle \bar{\psi}\psi \rangle ^{(1)}$
is always negative. By using the relation%
\begin{equation}
\int d\mathbf{k}_{p}\int_{\sqrt{\mathbf{k}_{p}^{2}+m_{\mathbf{n}_{q}}^{2}}%
}^{\infty }\frac{f(x)dx}{\sqrt{x^{2}-\mathbf{k}_{p}^{2}-m_{\mathbf{n}%
_{q}}^{2}}}=\frac{\pi ^{(p+1)/2}}{\Gamma ((p+1)/2)}\int_{m_{\mathbf{n}%
_{q}}}^{\infty }dx\,(x^{2}-m_{\mathbf{n}_{q}}^{2})^{(p-1)/2}f(x),
\label{rel2}
\end{equation}%
we find%
\begin{equation}
\langle \bar{\psi}\psi \rangle ^{(1)}=\frac{A_{p}N_{D}}{V_{q}}\sum_{\mathbf{n%
}_{q}\in \mathbf{Z}^{q}}\int_{m_{\mathbf{n}_{q}}}^{\infty }dx\,(x^{2}-m_{%
\mathbf{n}_{q}}^{2})^{(p-1)/2}(m-x)e^{-2xz},  \label{FC(1)1b}
\end{equation}%
with the notation
\begin{equation}
A_{p}=\frac{(4\pi )^{-(p+1)/2}}{\Gamma ((p+1)/2)}.  \label{Ap}
\end{equation}%
The integral in Eq.~(\ref{FC(1)1b}) is expressed in terms of the modified
Bessel function of the second type, namely%
\begin{equation}
\langle \bar{\psi}\psi \rangle ^{(1)}=\frac{N_{D}(2z)^{-p-1}}{(2\pi
)^{p/2+1}V_{q}}\sum_{\mathbf{n}_{q}\in \mathbf{Z}^{q}}\left[ 2mzf_{p/2}(2m_{%
\mathbf{n}_{q}}z)-f_{p/2+1}(2m_{\mathbf{n}_{q}}z)\right] .  \label{FC(1)2}
\end{equation}%
In the absence of compact dimensions, from Eq.~(\ref{FC(1)2}) one finds%
\begin{equation}
\langle \bar{\psi}\psi \rangle _{R^{D}}^{(1)}=\frac{N_{D}(2z)^{-D}}{(2\pi
)^{(D+1)/2}}\left[ 2mzf_{(D-1)/2}(2mz)-f_{(D+1)/2}(2mz)\right] .
\label{FC1q0}
\end{equation}

Let us consider some limiting cases of the general formula (\ref{FC(1)2}).
In the limit when the length of one of the compactified dimensions, say $%
z^{j}$, $j\geqslant p+2$, is large, $L_{j}\rightarrow \infty $, the dominant
contribution to the sum over $n_{j}$ in Eq.~(\ref{FC(1)2}) comes from large
values of $n_{j}$ and we can replace the corresponding summ by an integral,
with the help of the relation%
\begin{equation}
\frac{\pi }{L_{j}}\sum_{n_{j}=-\infty }^{+\infty }f(2\pi |n_{j}+\alpha
_{j}|/L_{j})\rightarrow \int_{0}^{\infty }dy\,f(y).  \label{IntForm}
\end{equation}%
The integral is evaluated by using the formula%
\begin{equation*}
\int_{0}^{\infty }dy\,f_{\nu }(c\sqrt{y^{2}+b^{2}})=\frac{1}{c}\sqrt{\frac{%
\pi }{2}}f_{\nu +1/2}(bc),
\end{equation*}%
and we can see that, from Eq.~(\ref{FC(1)2}), the corresponding formula is
obtained for the topology $R^{p+2}\times (S^{1})^{q-1}$. At small distances
from the boundary, $z\ll m^{-1},L_{l}$, the main contribution to the series
in Eq.~(\ref{FC(1)2}) comes from large values of $n_{l}$ and we can replace
the summation by the integration and to the leading order we find%
\begin{equation}
\langle \bar{\psi}\psi \rangle ^{(1)}\approx -\frac{N_{D}\Gamma ((D+1)/2)}{%
(4\pi )^{(D+1)/2}z^{D}}.  \label{FC1Near}
\end{equation}%
This leading behavior does not depend on the lengths of the compact
dimensions and, as it is seen from Eq.~(\ref{FC1q0}), coincides with
boundary induced part of the fermionic condensate for a single plate in a
space with trivial topology $R^{D}$ in the case of a massless field.

Now, let us consider the limit $L_{l}\ll z$. In this case, and for $\alpha
_{l}=0$, the main contribution comes from the zero mode with $\mathbf{n}%
_{q}=0$ and to the leading order we find%
\begin{equation*}
\langle \bar{\psi}\psi \rangle ^{(1)}\approx \frac{N(2z)^{-p-1}}{(2\pi
)^{p/2+1}V_{q}}\left[ 2mzf_{p/2}(2mz)-f_{p/2+1}(2mz)\right] .
\end{equation*}%
Comparing with (\ref{FC1q0}), we see that the quantity $V_{q}\langle \bar{%
\psi}\psi \rangle _{p,q}^{(1)}/N_{D}$ coincides with the corresponding
result for a plate in topologically trivial $(p+1)$-dimensional space, $%
R^{p+1}$. The contribution of the nonzero modes is exponentially suppressed
and, for $\alpha _{l}\neq 0$, the zero mode is absent. Assuming that $mz$ is
fixed, to leading order we have%
\begin{equation}
\langle \bar{\psi}\psi \rangle ^{(1)}=-\frac{N_{D}m_{0}^{p+1}e^{-2m_{0}z}}{%
2V_{q}(4\pi m_{0}z)^{(p+1)/2}},  \label{FC1Large}
\end{equation}%
where%
\begin{equation}
m_{0}^{2}=\sum_{l=p+2}^{D}(2\pi \alpha _{l}/L_{l})^{2}.  \label{m0}
\end{equation}%
In this case, the boundary induced part in the fermionic condensate is
exponentially suppressed.

Using (\ref{rel2}), we can also simplify the expression for the second plate
induced part in Eq.~(\ref{FC3}). Combining with Eq.~(\ref{FC(1)1b}) we find%
\begin{eqnarray}
\langle \bar{\psi}\psi \rangle &=&\langle \bar{\psi}\psi \rangle ^{(0)}-%
\frac{A_{p}N_{D}}{V_{q}}\sum_{\mathbf{n}_{q}\in \mathbf{Z}^{q}}\int_{m_{%
\mathbf{n}_{q}}}^{\infty }dx\,\frac{(x^{2}-m_{\mathbf{n}_{q}}^{2})^{(p-1)/2}%
}{\frac{x+m}{x-m}e^{2ax}+1}  \notag \\
&&\times \left[ (m+x)(e^{2xz}+e^{2ax-2xz})-2m\right] ,  \label{DeltaFC1}
\end{eqnarray}%
where the second term on the rhs is the boundary induced part, which is
always negative. For a massless field, by using the expansion $%
(e^{y}+1)^{-1}=-\sum_{n=1}^{\infty }(-1)^{n}e^{-ny}$, from this formula we
find
\begin{eqnarray}
\langle \bar{\psi}\psi \rangle &=&\langle \bar{\psi}\psi \rangle ^{(0)}+%
\frac{2N_{D}}{(4\pi )^{p/2+1}V_{q}}\sum_{\mathbf{n}_{q}\in \mathbf{Z}^{q}}k_{%
\mathbf{n}_{q}}^{p/2+1}\sum_{n=1}^{\infty }(-1)^{n}  \notag \\
&&\times \sum_{j=1,2}\frac{K_{p/2+1}(2k_{\mathbf{n}_{q}}(an-|a_{j}-z|))}{%
(an-|a_{j}-z|)^{p/2}},  \label{FCm0}
\end{eqnarray}%
where $a_{1}=0$ and $a_{2}=a$ and where $k_{\mathbf{n}_{q}}$ is defined by
Eq.~(\ref{mnq}). In the case of trivial topology $R^{D}$, from here we obtain%
\begin{equation}
\langle \bar{\psi}\psi \rangle _{R^{D}}=\frac{N_{D}\Gamma ((D+1)/2)}{(4\pi
)^{(D+1)/2}a^{D}}\sum_{n=1}^{\infty }(-1)^{n}\left[ \frac{1}{(n-z/a)^{D}}+%
\frac{1}{(n-1+z/a)^{D}}\right] .  \label{FCRDm0}
\end{equation}%
The series in these formula are given in terms of the Hurwitz zeta function.
The last expression can be further simplified, for odd values of $D$, as%
\begin{equation}
\langle \bar{\psi}\psi \rangle _{R^{D}}=-\frac{N_{D}\Gamma ((D+1)/2)}{%
2^{D+2}\pi ^{(D-1)/2}a^{D}}\frac{d^{D-1}}{dx^{D-1}}\frac{1}{\sin (\pi x)}%
\Big|_{x=z/a}.  \label{FCRDm0Odd}
\end{equation}%
In the special case $D=3$, from here we obtain the result given in Ref.~\cite%
{Lutk84}.

Extracting the parts corresponding to the single plates, the fermionic
condensate can also be presented in the form%
\begin{equation}
\langle \bar{\psi}\psi \rangle =\langle \bar{\psi}\psi \rangle
^{(0)}+\sum_{j=1,2}\langle \bar{\psi}\psi \rangle _{j}^{(1)}+\Delta \langle
\bar{\psi}\psi \rangle ,  \label{FCInterf}
\end{equation}%
where $\langle \bar{\psi}\psi \rangle _{j}^{(1)}$ is that part in the
condensate induced by a single plate located at $z=a_{j}$, with $a_{1}=0$
and $a_{2}=a$, and the interference term is given by the expression%
\begin{eqnarray}
\Delta \langle \bar{\psi}\psi \rangle &=&\frac{A_{p}N_{D}}{V_{q}}\sum_{%
\mathbf{n}_{q}\in \mathbf{Z}^{q}}\int_{m_{\mathbf{n}_{q}}}^{\infty }dx\,%
\frac{(x^{2}-m_{\mathbf{n}_{q}}^{2})^{(p-1)/2}}{\frac{x+m}{x-m}e^{2ax}+1}
\notag \\
&&\times \left[ 2m+(x-m)(e^{-2xz}+e^{2xz-2ax})\right] .  \label{FCInterf2}
\end{eqnarray}%
As it is seen from this formula, the interference part in the fermionic
condensate is always positive. Note that the divergences on the plates are
contained in the single plate parts and that the interference term is finite
for all values $0\leqslant z\leqslant a$. For a massless field, similar to
Eq.~(\ref{FCm0}), one finds the expression
\begin{equation}
\Delta \langle \bar{\psi}\psi \rangle =-\frac{2N_{D}}{(4\pi )^{p/2+1}V_{q}}%
\sum_{\mathbf{n}_{q}\in \mathbf{Z}^{q}}k_{\mathbf{n}_{q}}^{p/2+1}%
\sum_{n=1}^{\infty }(-1)^{n}\sum_{j=1,2}\frac{K_{p/2+1}(2k_{\mathbf{n}%
_{q}}(an+|a_{j}-z|))}{(an+|a_{j}-z|)^{p/2}}.  \label{DeltaFC1m0}
\end{equation}%
In the limit $L_{l}\ll a$ and for $\alpha _{l}=0$, the main contribution
comes from the zero mode with $\mathbf{n}_{q}=0$ and to leading order we
find $\Delta \langle \bar{\psi}\psi \rangle \approx N_{D}\Delta \langle \bar{%
\psi}\psi \rangle _{R^{p+1}}/(N_{p+1}V_{q})$, where%
\begin{eqnarray}
\Delta \langle \bar{\psi}\psi \rangle _{R^{p+1}}
&=&A_{p}N_{p+1}\int_{m}^{\infty }dx\,\frac{(x^{2}-m^{2})^{(p-1)/2}}{\frac{x+m%
}{x-m}e^{2ax}+1}  \notag \\
&&\times \left[ 2m+(x-m)(e^{-2xz}+e^{2xz-2ax})\right] ,  \label{FCRp+1}
\end{eqnarray}%
is the interference part in the fermionic condensate for the geometry of two
parallel plates in $(p+1)$-dimensional spacetime with trivial topology $%
R^{p+1}$. When $\alpha _{l}\neq 0$, there is no zero mode and the
interference part is exponentially suppressed by the factor $e^{-2am_{0}}$,
where $m_{0}$ is defined in Eq.~(\ref{m0}).

In the discussion above we have considered the fermionic condensate in the
region between the plates, $0<z<a$. For the regions $z<0$ and $z>a$, the
expression for the fermionic condensate has the form%
\begin{equation}
\langle \bar{\psi}\psi \rangle =\langle \bar{\psi}\psi \rangle
^{(0)}+\langle \bar{\psi}\psi \rangle ^{(1)},  \label{FCout}
\end{equation}%
where the plate-induced part is given by Eq.~(\ref{FC(1)2}) with the
replacement $z\rightarrow |z|$, in the region $z<0$, and with the
replacement $z\rightarrow z-a$, in the region $z>a$.

In the presence of a constant gauge field $A_{\mu }$, the corresponding
formulas for the fermionic condensate are obtained from those given above by
the replacement (\ref{PhaseRepl}). From these formulas it follows that the
fermionic condensate is a periodic function of $A_{l}L_{l}$ with the period
of the flux quantum $\Phi _{0}=2\pi /|e|$ ($\Phi _{0}=2\pi \hbar c/|e|$ in
standard units). It is an even function of $\tilde{\alpha}_{l}$.

\begin{figure}[tbph]
\begin{center}
\begin{tabular}{cc}
\epsfig{figure=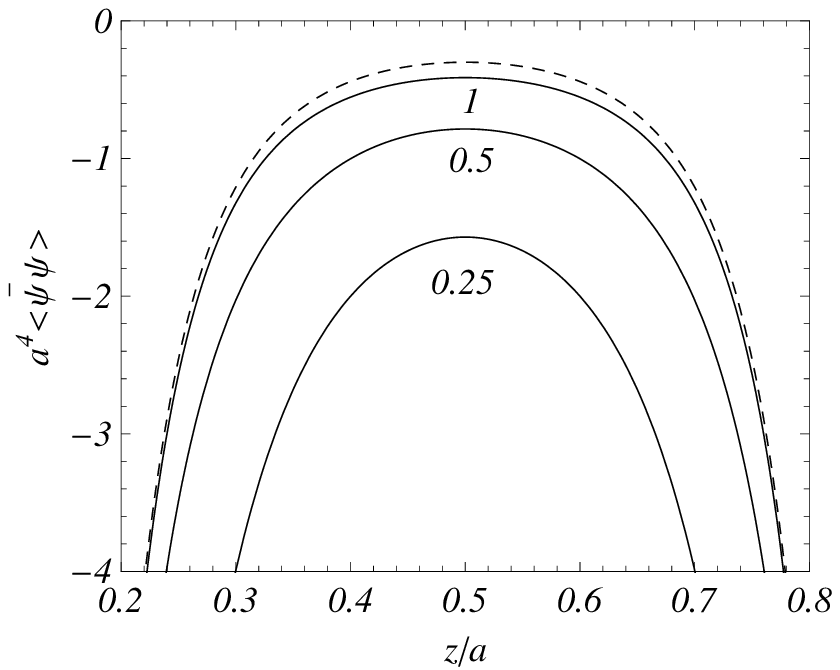,width=7.cm,height=6cm} & \quad %
\epsfig{figure=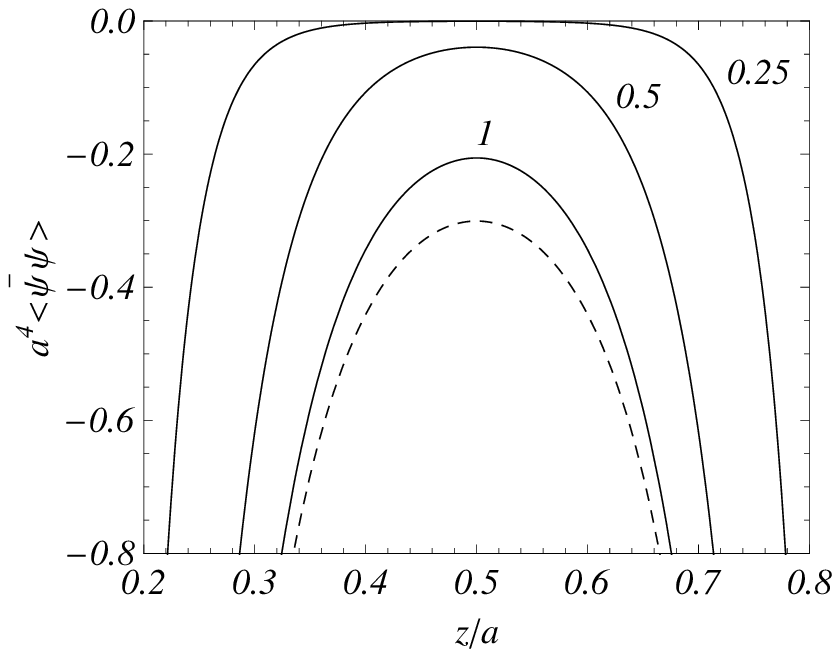,width=7.cm,height=6cm}%
\end{tabular}%
\end{center}
\caption{Fermionic condensate between two plates, in the model with spatial
topology $R^{3}\times S^{1}$, as a function of $z/a$ for untwisted ($\protect%
\alpha _{4}=0$, left plot) and twisted ($\protect\alpha _{4}=1/2$, right
plot) fields. The numbers near the curves correspond to the values of the
ratio $L_{4}/a$. The dashed lines in both plots correspond to the fermionic
condensate for two plates on background of the space with trivial topology $%
R^{4}$ [see Eq.~(\protect\ref{FCRDm0})].}
\label{fig1}
\end{figure}

The formulas corresponding to the special case with a single compact
dimension are obtained from those given above by taking $p=D-2$, $q=1$, and $%
k_{\mathbf{n}_{q}}=2\pi |n_{D}+\alpha _{D}|/L_{D}$. In Fig.~\ref{fig1} we
plot, for the simplest Kaluza-Klein-type model with spatial topology $%
R^{3}\times S^{1}$, the dependence of the fermionic condensate for a
massless fermionic field vs the ratio $z/a$, in the region between the
plates, for untwisted and twisted fermionic fields, for several values of
the ratio $L/a$. The dashed lines correspond to the fermionic condensate for
the geometry of two plates in a space with topology $R^{4}$, defined by Eq.~(%
\ref{FCRDm0}). Note that, for a massless field, the pure topological part
vanishes.
As it is seen from the graphs, for an untwisted (twisted) field the absolute
value of the fermionic condensate increases with decreasing (increasing)
length of the compact dimension.

\section{Vacuum expectation value of the energy-momentum tensor}

\label{sec:EMT}

Another important local characteristic of the fermionic vacuum is the VEV of
the energy-momentum tensor. In order to find this we use the mode-sum
formula
\begin{equation}
\langle T_{\mu \nu }\rangle =\frac{i}{2}\sum_{\beta }[\bar{\psi}_{\beta
}^{(-)}\gamma _{(\mu }\partial _{\nu )}\psi _{\beta }^{(-)}-(\partial _{(\mu
}\bar{\psi}_{\beta }^{(-)})\gamma _{\nu )}\psi _{\beta }^{(-)}]\ ,
\label{modesum}
\end{equation}%
where the brackets denote the symmetrization over the indices enclosed. By
taking into account the expressions for the spinor eigenfunctions, we find
the following expressions for the components of the vacuum energy-momentum
tensor (no summation over $\mu $)
\begin{equation}
\langle T_{\mu }^{\nu }\rangle =-\frac{N_{D}\delta _{\mu }^{\nu }}{2aV_{q}}%
\sum_{\mathbf{n}_{q}\in \mathbf{Z}^{q}}\int \frac{d\mathbf{k}_{p}}{(2\pi
)^{p}}\sum_{n=1}^{\infty }\frac{f^{(\mu )}(\lambda _{n})/\omega }{1-\sin
(2\lambda _{n})/(2\lambda _{n})},  \label{EMT}
\end{equation}%
where $\omega $ is defined in Eq.~(\ref{omeg}) and%
\begin{eqnarray}
f^{(0)}(x) &=&\omega ^{2}\left[ 1-\cos \left( x\right) \cos \left(
x(2z^{p+1}/a-1)\right) \right] ,  \notag \\
f^{(l)}(x) &=&-(k_{l}^{2}/\omega ^{2})f^{(0)}(x),\;l\neq 0,p+1,  \label{fl}
\\
f^{(p+1)}(x) &=&-k_{p+1}^{2}.  \notag
\end{eqnarray}%
As in the case of the fermionic condensate, we assume the presence of the
cutoff function in (\ref{EMT}). It can be checked that the VEVs of the
separate components obey the trace relation%
\begin{equation}
\langle T_{\mu }^{\mu }\rangle =m\langle \bar{\psi}\psi \rangle ,
\label{TraceRel}
\end{equation}%
with the fermionic condensate given by Eq.~(\ref{FC2}).

The total vacuum energy (per unit volume along the directions $z_{l}$, $%
l=1,2,\ldots ,p$) in the region $0<z_{p+1}<a$, $0\leqslant z_{l}\leqslant
L_{l}$, $l=p+2,\ldots ,D$, is obtained integrating $\langle T_{0}^{0}\rangle
$ over this region. By taking into account the expression for the energy
density from Eq.~(\ref{EMT}) and Eq.~(\ref{kpvalues}) for the eigenvalues,
we obtain the following expression for the vacuum energy%
\begin{equation}
E=-\frac{N_{D}}{2}\sum_{\mathbf{n}_{q}\in \mathbf{Z}^{q}}\int \frac{d\mathbf{%
k}_{p}}{(2\pi )^{p}}\sum_{n=1}^{\infty }\sqrt{\mathbf{k}_{p}^{2}+k_{\mathbf{n%
}_{q}}^{2}+\lambda _{n}^{2}/a^{2}+m^{2}}.  \label{Toten}
\end{equation}%
It coincides with the total energy of the fermionic vacuum in the region
under consideration, evaluated as the sum of zero-point energies of
elementary oscillators. Hence, we have shown that the vacuum energy obtained
by integration of the energy density in the region between the plates does
coincide with the energy evaluated as the sum of the zero-point energies of
elementary oscillators. This means that the surface energy located on the
boundaries is zero for the boundary conditions under consideration. Note
that the surface energy vanishes for scalar fields with Dirichlet or Neumann
boundary conditions as well, but this is not the case for a scalar field
with Robin boundary conditions (see Refs.~\cite{Rome02,Saha04}). The total
vacuum energy in the region between the plates, defined by Eq.~(\ref{Toten}%
), is investigated in Ref.~\cite{Bell09}. Here we will be concerned with the
vacuum energy density and stresses.

For the evaluation of the VEVs, given by Eq.~(\ref{EMT}), we apply to the
sums over $n$ the summation formula (\ref{Abel-Plan}). After steps similar
to those already described in the case of the fermionic condensate, the VEV
of the energy-momentum tensor can be expressed in the decomposed form (no
summation over $\mu $)%
\begin{eqnarray}
\langle T_{\mu }^{\nu }\rangle  &=&\langle T_{\mu }^{\nu }\rangle
^{(0)}+\langle T_{\mu }^{\nu }\rangle ^{(1)}-\frac{A_{p}N_{D}}{V_{q}}\delta
_{\mu }^{\nu }  \notag \\
&&\times \sum_{\mathbf{n}_{q}\in \mathbf{Z}^{q}}\int_{m_{\mathbf{n}%
_{q}}}^{\infty }dx\frac{(x^{2}-m_{\mathbf{n}_{q}}^{2})^{(p-1)/2}}{\frac{x+m}{%
x-m}e^{2ax}+1}F_{\mu }(x),  \label{EMTdec}
\end{eqnarray}%
where $A_{p}$ is given by Eq.~(\ref{Ap}) and we have defined the functions%
\begin{eqnarray}
F_{\mu }(x) &=&\frac{x^{2}-m_{\mathbf{n}_{q}}^{2}}{p+1}\left( 2+\frac{%
me^{2xz}}{x-m}-\frac{me^{-2xz}}{x+m}\right) ,\;\mu =0,\ldots ,p,  \notag \\
F_{\mu }(x) &=&k_{\mu }^{2}\left( 2+\frac{me^{2xz}}{x-m}-\frac{me^{-2xz}}{x+m%
}\right) ,\;\mu =p+2,\ldots ,D,  \label{Fl} \\
F_{\mu }(x) &=&-2x^{2},\;\mu =p+1.  \notag
\end{eqnarray}%
In Eq.~(\ref{EMTdec}), the term (no summation over $\mu $)%
\begin{equation}
\langle T_{\mu }^{\nu }\rangle ^{(0)}=-\frac{N_{D}\delta _{\mu }^{\nu }}{%
2V_{q}}\sum_{\mathbf{n}_{q}\in \mathbf{Z}^{q}}\int \frac{d\mathbf{k}_{p+1}}{%
(2\pi )^{p+1}}\frac{f_{0}^{(\mu )}}{\sqrt{\mathbf{k}_{p+1}^{2}+m_{\mathbf{n}%
_{q}}^{2}}},  \label{EMT0}
\end{equation}%
with $f_{0}^{(0)}=\mathbf{k}_{p+1}^{2}+m_{\mathbf{n}_{q}}^{2}$, $%
f_{0}^{(l)}=-k_{l}^{2}$, $l=1,\ldots ,p+1$, is the VEV\ of the
energy-momentum tensor for the topology $R^{p+1}\times (S^{1})^{q}$ when the
boundaries are absent. This background is homogeneous and the corresponding
densities are uniform. The second term on the rhs of Eq.~(\ref{EMTdec}) is
induced by the plate at $z=0$ when the second plate is absent (again no
summation over $\mu $):%
\begin{equation}
\langle T_{\mu }^{\nu }\rangle ^{(1)}=\frac{N_{D}m\delta _{\mu }^{\nu }}{%
4\pi V_{q}}\int \frac{d\mathbf{k}_{p}}{(2\pi )^{p}}\sum_{\mathbf{n}_{q}\in
\mathbf{Z}^{q}}\int_{0}^{\infty }dx\frac{f_{0}^{(\mu )}(x)}{\sqrt{x^{2}+%
\mathbf{k}_{p}^{2}+m_{\mathbf{n}_{q}}^{2}}}\left( \frac{e^{-2ixz}}{m+ix}+%
\frac{e^{2ixz}}{m-ix}\right) ,  \label{EMT1pl}
\end{equation}%
where%
\begin{eqnarray}
f_{0}^{(0)}(x) &=&x^{2}+\mathbf{k}_{p}^{2}+m_{\mathbf{n}_{q}}^{2},%
\;f_{0}^{(p+1)}(x)=0,  \notag \\
f_{0}^{(l)}(x) &=&-k_{l}^{2},\;l=1,\ldots ,p,p+2,\ldots ,D.  \label{f0}
\end{eqnarray}%
The last term on the rhs of Eq.~(\ref{EMTdec}) is induced by the presence of
the second plate. Note that the vacuum stress along the direction normal to
the plates vanishes in the geometry of a single plate and is uniform in the
region between the two plates. The renormalized expressions for the purely
topological part $\langle T_{\mu }^{\nu }\rangle ^{(0)}$ are given in Ref.~%
\cite{Bell09b}. In particular, for the corresponding energy density one has%
\begin{eqnarray}
&&\langle T_{0}^{0}\rangle ^{(0)}=\frac{N_{D}}{(2\pi )^{(D+1)/2}}%
\sideset{}{'}{\sum}_{\mathbf{n}_{q}\in \mathbf{Z}^{q}}\cos (2\pi \mathbf{n}%
_{q}\cdot \widetilde{\boldsymbol{\alpha }}_{q})  \notag \\
&&\qquad \times \frac{f_{(D+1)/2}(m\sqrt{L_{p+2}^{2}n_{p+2}^{2}+\cdots
+L_{D}^{2}n_{D}^{2}})}{(L_{p+2}^{2}n_{p+2}^{2}+\cdots
+L_{D}^{2}n_{D}^{2})^{(D+1)/2}}.  \label{Tp0}
\end{eqnarray}%
where, as before, the prime means that the term $\mathbf{n}_{q}=0$ is
excluded from the sum. Note that (no summation over $\mu $) $\langle T_{\mu
}^{\mu }\rangle ^{(0)}=\langle T_{0}^{0}\rangle ^{(0)}$ for $\mu =1,\ldots
,p+1$. In the discussion below we will be concerned with the boundary
induced parts.

The single plate part, Eq.~(\ref{EMT1pl}), can be further simplified. With
this aim, in the term with the exponent $e^{2ixz}$ ($e^{-2ixz}$), we rotate
the integration contour in the complex plane $x$ by the angle $\pi /2$ ($%
-\pi /2$). By using the integration formula (\ref{rel2}), we find the
following result (no summation over $\mu $)%
\begin{equation}
\langle T_{\mu }^{\mu }\rangle ^{(1)}=-m\frac{A_{p}N_{D}}{V_{q}}\sum_{%
\mathbf{n}_{q}\in \mathbf{Z}^{q}}\int_{m_{\mathbf{n}_{q}}}^{\infty }dx\,%
\frac{(x^{2}-m_{\mathbf{n}_{q}}^{2})^{(p-1)/2}}{(x+m)e^{2xz}}F_{0}^{(\mu
)}(x),  \label{EMT1pl1}
\end{equation}%
with the notations
\begin{eqnarray}
F_{0}^{(\mu )}(x) &=&\frac{x^{2}-m_{\mathbf{n}_{q}}^{2}}{p+1},\;\mu
=0,1,\ldots ,p,  \notag \\
F_{0}^{(\mu l)}(x) &=&k_{\mu }^{2},\;\mu =p+2,\ldots ,D,  \label{F0l}
\end{eqnarray}%
and $F_{0}^{(p+1)}(x)=0$. Note that the boundary induced parts in the vacuum
stresses along the uncompactified directions parallel to the plates are
equal to the boundary induced part in the energy density. This result is a
consequence of the Lorentz invariance of the problem along these directions.
The energy density is negative everywhere. In the absence of compact
dimensions, from Eq.~(\ref{EMT1pl1}) one finds (no summation over $\mu $)
\begin{equation}
\langle T_{\mu }^{\mu }\rangle ^{(1)}=m\langle \bar{\psi}\psi \rangle
^{(1)}/D,  \label{EMT1plRD}
\end{equation}%
for $\mu =0,\ldots D-1$, and $\langle T_{D}^{D}\rangle ^{(1)}=0$. In Eq.~(%
\ref{EMT1plRD}) $\langle \bar{\psi}\psi \rangle ^{(1)}$ is given by Eq.~(\ref%
{FC1q0}). Of course, Eq.~(\ref{EMT1plRD}) is a direct consequence of the
trace relation (\ref{TraceRel}).

When the length of the $j$-th compact dimension is large, we replace the
summation over $n_{j}$ in Eq.~(\ref{EMT1pl1}) by an integral, with the help
of Eq.~(\ref{IntForm}). As next step, we introduce a new integration
variable $v=\sqrt{x^{2}-m_{\mathbf{n}_{q}}^{2}}$. After changing to polar
coordinates in the $(y,v)$-plane, the angular part of the integral is
evaluated explicitly and, as it can be seen from Eq.~(\ref{EMT1pl1}), to
leading order the result is obtained for the topology $R^{p+2}\times
(S^{1})^{q-1}$.

For points near the plate, $z\ll L_{l}$, the dominant contribution to the
series in Eq.~(\ref{EMT1pl1}) comes from large values of $n_{l}$. Replacing
the corresponding summations by integrations, we see that, to leading order,
the behavior of the VEV coincides with the one for topologically trivial
space. If, in addition, $z\ll m^{-1}$, combining Eq.~(\ref{EMT1plRD}) with
Eq.~(\ref{FC1Near}), one finds (no summation over $\mu $)%
\begin{equation}
\langle T_{\mu }^{\mu }\rangle ^{(1)}\approx -\frac{mN_{D}\Gamma ((D+1)/2)}{%
(4\pi )^{(D+1)/2}Dz^{D}},  \label{EMT1plNear}
\end{equation}%
for $\mu =0,\ldots ,D-1$. For $L_{l}\ll z$ the behavior of the VEVs is
essentially different, depending on the phases $\alpha _{l}$ in the
periodicity conditions. When $\alpha _{l}=0$, $l=0,\ldots ,D$, the
contribution of the nonzero modes is exponentially suppressed and the
dominant contribution comes from the zero mode $\mathbf{n}_{q}=0$. For the
components along the uncompactified dimensions, to leading order we have (no
summation over $\mu $) $\langle T_{\mu }^{\mu }\rangle ^{(1)}\approx
N_{D}\langle T_{\mu }^{\mu }\rangle _{R^{p+1}}^{(1)}/(V_{q}N_{p+1})$, $\mu
=0,\ldots ,p+1$, where $\langle T_{\mu }^{\nu }\rangle _{R^{p+1}}^{(1)}$ is
the corresponding VEV for two parallel plates in the space $R^{p+1}$ located
at $z=0$ and $z=a$. For the stress along the $l$-th compact dimension the
contribution of the zero mode vanishes and, in the case under consideration (%
$\alpha _{l}=0$), the dominant contribution comes from the modes with $%
n_{l}=\pm 1$, $n_{j}=0$, $j\neq l$, and the stress is exponentially
suppressed (no summation over $l$): $\langle T_{l}^{l}\rangle ^{(1)}\propto
e^{-2z/L_{l}}$. We have also assumed that $L_{l}\ll m^{-1}$. For $\alpha
_{l}\neq 0$ the VEVs are exponentially suppressed and to leading order we
have (no summation over $\mu $)%
\begin{equation}
\langle T_{\mu }^{\mu }\rangle ^{(1)}\approx -\frac{(4\pi
)^{-(p+1)/2}N_{D}mm_{0}{}^{p+1}}{4V_{q}(m_{0}z)^{(p+3)/2}e^{2m_{0}z}},
\label{EMT1plLarg}
\end{equation}%
for $\mu =0,\ldots ,p$ and $m_{0}$ one recovers Eq.~(\ref{m0}). We have a
similar exponential suppression for the stresses along the compact
dimensions.

Now we return to the two plate geometry. Combining Eq.~(\ref{EMT1pl1}) with
Eq.~(\ref{EMTdec}), for the VEV of the energy-momentum tensor in the region
between the plates, we find (no summation over $\mu $)%
\begin{equation}
\langle T_{\mu }^{\nu }\rangle =\langle T_{\mu }^{\nu }\rangle ^{(0)}-\delta
_{\mu }^{\nu }\frac{A_{p}N_{D}}{V_{q}}\sum_{\mathbf{n}_{q}\in \mathbf{Z}%
^{q}}\int_{m_{\mathbf{n}_{q}}}^{\infty }dx\,\frac{(x^{2}-m_{\mathbf{n}%
_{q}}^{2})^{(p-1)/2}}{\frac{x+m}{x-m}e^{2ax}+1}G_{\mu }(x),  \label{EMT2pl}
\end{equation}%
where%
\begin{eqnarray}
G_{\mu }(x) &=&\frac{x^{2}-m_{\mathbf{n}_{q}}^{2}}{p+1}\left[ 2+\frac{m}{x-m}%
(e^{2xz}+e^{2ax-2xz})\right] ,\;\mu =0,1,\ldots ,p,  \label{Gl1} \\
G_{\mu }(x) &=&k_{\mu }^{2}\left[ 2+\frac{m}{x-m}(e^{2xz}+e^{2ax-2xz})\right]
,\;\mu =p+2,\ldots ,D,  \label{Gl2}
\end{eqnarray}%
and $G_{p+1}(x)=F_{p+1}(x)$. From here, it follows that the boundary induced
part in the vacuum energy density is always negative. It can be easily
checked that the boundary induced parts obey the trace relation (\ref%
{TraceRel}). For the VEV in the geometry of two parallel plates located at $%
z=0$ and $z=a$ in a $D$-dimensional space with topology $R^{D}$, one has (no
summation over $\mu $)%
\begin{equation}
\langle T_{\mu }^{\nu }\rangle _{R^{D}}=-\frac{N_{D}\delta _{\mu }^{\nu }}{%
(4\pi )^{D/2}\Gamma (D/2)}\int_{m}^{\infty }dx\,\frac{(x^{2}-m^{2})^{(p-1)/2}%
}{\frac{x+m}{x-m}e^{2ax}+1}G_{0\mu }(x),  \label{EMTRD}
\end{equation}%
where the expression for $G_{0\mu }(x)$, with $\mu =0,1,\ldots ,D-1$, is
obtained from Eq.~(\ref{Gl1}) by the replacement $m_{\mathbf{n}%
_{q}}\rightarrow m$ and $G_{0D}(x)=-2x^{2}$. For a massless field this
result reduces to%
\begin{equation}
\langle T_{\mu }^{\nu }\rangle _{R^{D}}=-\frac{(1-2^{-D})N_{D}}{(4\pi
)^{(D+1)/2}a^{D+1}}\zeta (D+1)\Gamma ((D+1)/2)\text{diag}(1,\ldots ,1,-D).
\label{EMTRDm0}
\end{equation}%
Note that $\langle T_{\mu }^{\nu }\rangle _{R^{D}}=(1-2^{-D})N_{D}\langle
T_{\mu }^{\nu }\rangle _{R^{D}}^{\text{(sc)}}$, where $\langle T_{\mu }^{\nu
}\rangle _{R^{D}}^{\text{(sc)}}$ is the corresponding VEV for a scalar field
with Dirichlet or Neumann boundary conditions on the plates.

For a massless field and in the presence of compact dimensions, the
expressions for the VEV of the energy-momentum tensor are obtained from
general formulas (\ref{EMT2pl}) by putting $m=0$. Alternative expressions
are derived using the expansion $(e^{y}+1)^{-1}=-\sum_{n=1}^{\infty
}(-1)^{n}e^{-ny}$. After integration one finds (no summation over $\mu $)%
\begin{equation}
\langle T_{\mu }^{\mu }\rangle =\langle T_{\mu }^{\mu }\rangle ^{(0)}+\frac{%
2N_{D}}{(2\pi )^{p/2+1}V_{q}}\sum_{n=1}^{\infty }\frac{(-1)^{n}}{(2an)^{p+2}}%
\sum_{\mathbf{n}_{q}\in \mathbf{Z}^{q}}G_{\mu }^{(0)}(2ank_{\mathbf{n}_{q}}),
\label{DeltaEMTm0}
\end{equation}%
with the notations%
\begin{eqnarray}
G_{\mu }^{(0)}(x) &=&f_{p/2+1}(x),\;\mu =0,\ldots ,p,  \notag \\
G_{\mu }^{(0)}(x) &=&(2ank_{\mu })^{2}f_{p/2}(x),\;\mu =p+2,\ldots ,D,
\label{Hl0} \\
G_{p+1}^{(0)}(x) &=&-(p+1)f_{p/2+1}(x)-x^{2}f_{p/2}(x).  \notag
\end{eqnarray}%
The corresponding vacuum densities are uniform. In this case the boundary
induced part in the total energy (per unit volume along uncompactified
dimensions) of the vacuum is $aV_{q}\Delta \langle T_{0}^{0}\rangle $ and
the corresponding result obtained from Eq.~(\ref{DeltaEMTm0}) coincides with
the result derived in Ref.~\cite{Bell09} by using zeta function techniques.

Similar to Eq.~(\ref{FCInterf}), the VEV of the energy-momentum tensor may
be presented in the decomposed form%
\begin{equation}
\langle T_{\mu }^{\nu }\rangle =\langle T_{\mu }^{\nu }\rangle
^{(0)}+\sum_{j=1,2}\langle T_{\mu }^{\nu }\rangle _{j}^{(1)}+\Delta \langle
T_{\mu }^{\nu }\rangle ,  \label{EMTDec}
\end{equation}%
where $\langle T_{\mu }^{\nu }\rangle _{j}^{(1)}$ is the part in the VEV\
induced by a single plate at $z=a_{j}$, with $a_{1}=0$, $a_{2}=0$. The
interference part in Eq.~(\ref{EMTdec}) is given by the expression (no
summation over $\mu $)%
\begin{equation}
\Delta \langle T_{\mu }^{\nu }\rangle =-\delta _{\mu }^{\nu }\frac{A_{p}N_{D}%
}{V_{q}}\sum_{\mathbf{n}_{q}\in \mathbf{Z}^{q}}\int_{m_{\mathbf{n}%
_{q}}}^{\infty }dx\frac{(x^{2}-m_{\mathbf{n}_{q}}^{2})^{(p-1)/2}}{\frac{x+m}{%
x-m}e^{2ax}+1}H_{\mu }(x),  \label{DeltEMT}
\end{equation}%
with the notation%
\begin{eqnarray}
H_{\mu }(x) &=&\frac{x^{2}-m_{\mathbf{n}_{q}}^{2}}{p+1}\left[ 2-\frac{m}{x+m}%
\left( e^{-2xz}+e^{-2x(a-z)}\right) \right] ,\;\mu =0,1,\ldots ,p,  \notag \\
H_{\mu }(x) &=&k_{\mu }^{2}\left[ 2-\frac{m}{x+m}\left(
e^{-2xz}+e^{-2x(a-z)}\right) \right] ,\;\mu =p+2,\ldots ,D,  \label{Hl}
\end{eqnarray}%
and $H_{p+1}(x)=F_{p+1}(x)$. The surface divergences in the VEV of the
energy-momentum tensor are contained in the single plate parts only and the
interference part is everywhere finite. For a massless field, the single
plate parts in the VEV of the energy-momentum tensor vanish and the
interference part coincides with the second term on the rhs of Eq.~(\ref%
{DeltaEMTm0}). In the limit $L_{l}\ll a$ and for $\alpha _{l}=0$ the
dominant contribution to the interference part comes from the zero mode $%
\mathbf{n}_{q}$. To leading order, we find (no summation over $\mu $) $%
\Delta \langle T_{\mu }^{\mu }\rangle =N_{D}\Delta \langle T_{\mu }^{\mu
}\rangle _{R^{p+1}}/(V_{q}N_{p+1})$, $\mu =0,1,\ldots ,p+1$, where $\Delta
\langle T_{\mu }^{\mu }\rangle _{R^{p+1}}$ is the VEV for two plates in a $%
(p+1)$-dimensional spacetime with topology $R^{p+1}$. For the stress along
the $l$-th compact dimension, the contribution of the zero mode vanishes.
The dominant contribution comes from the modes with $n_{l}=\pm 1$, $n_{j}=0$%
, $j\neq l$, and the stress is exponentially suppressed: $\Delta \langle
T_{l}^{l}\rangle \propto e^{-2a/L_{l}}$, where we have additionally assumed
that $L_{l}\ll m^{-1}$. For $\alpha _{l}\neq 0$ and $L_{l}\ll a$ the
interference part in the VEV of the energy-momentum tensor is suppressed by
the factor $e^{-2am_{0}}$, with $m_{0}$ defined in Eq.~(\ref{m0}).

In the regions $z^{p+1}<0$ and $z^{p+1}>a$ the expression for the VEV of the
energy-momentum tensor has the form%
\begin{equation}
\langle T_{\mu }^{\nu }\rangle =\langle T_{\mu }^{\nu }\rangle
^{(0)}+\langle T_{\mu }^{\nu }\rangle ^{(1)},  \label{EMT1b}
\end{equation}%
where the boundary induced part is given by Eq.~(\ref{EMT1pl1}), with the
replacements $z\rightarrow |z|$ and $z\rightarrow z-a$ for the regions $z<0$
and $z>a$, respectively. In the presence of a constant gauge field $A_{\mu }$%
, the formulas for the VEV of the energy-momentum tensor are obtained by the
replacement (\ref{PhaseRepl}). As in the case of the fermionic condensate,
the vacuum energy-momentum tensor is a periodic function of $A_{l}L_{l}$
with the period of the flux quantum.

The vacuum forces per unit surface of the plates are equal to the normal
stress evaluated at $z=0$ and $z=a$. By taking into account that $\langle
T_{p+1}^{p+1}\rangle ^{(1)}=0$ one finds, in the region between the plates,%
\begin{eqnarray}
&&P(0+)=P(a-)=-\langle T_{p+1}^{p+1}\rangle ^{(0)}-\Delta \langle
T_{p+1}^{p+1}\rangle  \notag \\
&&\qquad =-\langle T_{p+1}^{p+1}\rangle ^{(0)}-\frac{2A_{p}N_{D}}{V_{q}}%
\sum_{\mathbf{n}_{q}\in \mathbf{Z}^{q}}\int_{\widetilde{m}_{\mathbf{n}%
_{q}}}^{\infty }dx\frac{x^{2}(x^{2}-\widetilde{m}_{\mathbf{n}%
_{q}}^{2})^{(p-1)/2}}{\frac{x+m}{x-m}e^{2ax}+1},  \label{P}
\end{eqnarray}%
where $\widetilde{m}_{\mathbf{n}_{q}}^{2}=\sum_{l=p+2}^{D}[2\pi (n_{l}+%
\tilde{\alpha}_{l})/L_{l}]^{2}+m^{2}$. The term $-\langle
T_{p+1}^{p+1}\rangle ^{(0)}$ does not depend on the plate separation, it is
a purely topological part in the vacuum forces. For this term, one has $%
\langle T_{p+1}^{p+1}\rangle ^{(0)}=\langle T_{0}^{0}\rangle ^{(0)}$, with $%
\langle T_{0}^{0}\rangle ^{(0)}$ given by Eq.~(\ref{Tp0}). The last term on
the rhs of Eq.~(\ref{P}) is induced by the presence of the second plate and
determines the interaction forces between the plates. This term is negative
and the interaction forces between the plates are always attractive, with
independence of the periodicity conditions along compact dimensions and of
the value of the gauge potential. In absence of the gauge field, Eq.~(\ref{P}%
) coincides with the result obtained in Ref.~\cite{Bell09}, by
differentiation of the total Casimir energy. For the vacuum forces in the
regions $z<0$ and $z>a$, only the pure topological part contributes, and one
has%
\begin{equation}
P(0-)=P(a+)=-\langle T_{p+1}^{p+1}\rangle ^{(0)}.  \label{Pext}
\end{equation}%
When the quantum field lives in all regions, the pure topological parts of
the force acting from the left and from the right-hand sides of the plate
compensate and the resulting force is determined by the last term on the rhs
of Eq.~(\ref{P}). In some important physical situations (bag model in QCD,
finite-length carbon nanotubes, higher-dimensional models with orbifolded
extra dimensions) the quantum field is confined to the interior of some
region and there is no field outside. For the problem under consideration,
if the quantum field is confined in the region between the plates, the total
Casimir force, acting per unit surface of the plate, is determined by Eq.~(%
\ref{P}) and the pure topological part contributes as well. The resulting
force can be either attractive or repulsive, depending on the phases in the
periodicity conditions along the compact dimensions, and also on the gauge
potential. In this case, the Casimir effect could be used as a stabilization
mechanism for both the interplate distance and the size of the compact
subspace in Kaluza-Klein-type models and in braneworlds. This is a quite
remarkable and very useful result, because this force could be, in
principle, easily controlled.

\begin{figure}[tbph]
\begin{center}
\epsfig{figure=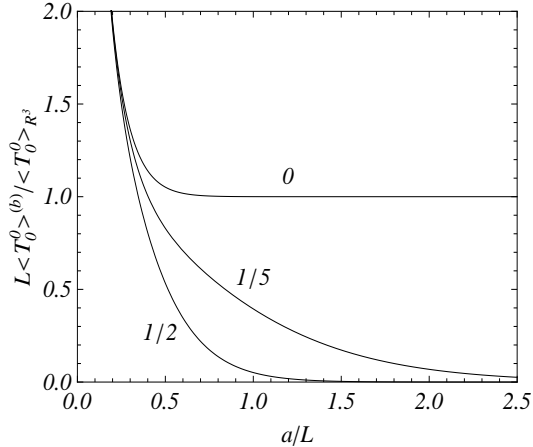,width=7.cm,height=6.cm}
\end{center}
\caption{Ratio of the boundary induced part in the Casimir energy for two
parallel plates in the spacetime with spatial topology $R^{3}\times S^{1}$
by the standard Casimir energy in $R^{3}$, for a massless fermionic field,
as a function of $a/L$. The values on each of the curves correspond to those
of the parameter $\protect\alpha _{4}$.}
\label{fig2}
\end{figure}

In Fig.~\ref{fig2}, we have plotted the ratio of the boundary
induced part in the Casimir energy (per unit surface along
uncompactified dimensions) for two parallel plates in the
spacetime with topology $R^{3}\times S^{1}$ by
the Casimir energy in $R^{3}$, $E_{R^{3}\times S^{1}}^{\text{(b)}%
}/E_{R^{3}}=L\langle T_{0}^{0}\rangle ^{\text{(b)}}/\langle T_{0}^{0}\rangle
_{R^{3}}$ ($L$ being the length of the compact dimension), with  $\langle
T_{0}^{0}\rangle ^{\text{(b)}}=\langle T_{0}^{0}\rangle -\langle
T_{0}^{0}\rangle ^{(0)}$, for a massless fermionic field, as a function of $%
a/L$. The values on each of the curves correspond to those of the parameter $%
\alpha _{4}$. Note that from Eq.~(\ref{EMTRDm0}) one has $\langle
T_{0}^{0}\rangle _{R^{3}}=-7\pi ^{2}/(2880a^{4})$. The feature described
before is apparent: for large values of $a/L$ the Casimir energy is
suppressed for $\alpha _{4}\neq 0$.

For a massless case the fermionic field is conformally invariant. We can
generate the corresponding VEVs in conformally flat spacetimes by using
standard conformal transformation techniques (see, for instance, \cite%
{Birr82}) and the formulas given above. Consider a conformally-flat
spacetime with the line-element%
\begin{equation}
ds^{2}=\Omega ^{2}(z_{l})(dt^{2}-\sum_{i=1}^{D}(dz_{i})^{2}),  \label{ds2CF}
\end{equation}%
and spatial topology $R^{p+1}\times (S^{1})^{q}$, with $0\leqslant
z_{l}\leqslant L_{l}$, $l=p+2,\ldots ,D$. As before, we assume the presence
of two boundaries located at $z=0$ and $z=a$ ($z_{p+1}\equiv z$), with the
boundary conditions $[1+i\gamma _{(\Omega )}^{\mu }n_{(\Omega )\mu }]\psi
_{(\Omega )}=0$, where the subscript $\Omega $ specifies the quantities on
the background described by Eq.~(\ref{ds2CF}). For curved space gamma
matrices we have $\gamma _{(\Omega )}^{\mu }=e_{l}^{\mu }\gamma ^{l}$, with
the tetrad field $e_{l}^{\mu }=\Omega ^{-1}\delta _{l}^{\mu }$. By taking
into account that, under the conformal transformation for a massless field,
one has $\psi _{(\Omega )}=\Omega ^{-D}\psi $ and $n_{(\Omega )\mu }=\Omega
n_{\mu }$, we see that the MIT bag boundary condition is conformally
invariant. For the fermionic condensate and the VEV of the energy-momentum
tensor on the spacetime (\ref{ds2CF}), one finds%
\begin{eqnarray}
\langle \bar{\psi}\psi \rangle _{(\Omega )} &=&\langle \bar{\psi}\psi
\rangle _{(\Omega ),R^{D}}+\Omega ^{-D}\langle \bar{\psi}\psi \rangle ,
\notag \\
\langle T_{\mu }^{\nu }\rangle _{(\Omega )} &=&\langle T_{\mu }^{\nu
}\rangle _{(\Omega ),R^{D}}+\Omega ^{-D-1}\langle T_{\mu }^{\nu }\rangle ,
\label{FCTilCF}
\end{eqnarray}%
where $\langle \bar{\psi}\psi \rangle _{(\Omega ),R^{D}}$ and $\langle
T_{\mu }^{\nu }\rangle _{(\Omega ),R^{D}}$ are the corresponding VEVs for
spacetime (\ref{ds2CF}) with trivial spatial topology $R^{D}$. Note that for
points away from the boundaries, all divergences are contained in these
terms and the renormalization procedure is needed for them only. The second
terms on the rhs of Eq.~(\ref{FCTilCF}) are induced by the non-trivial
topology and by the boundaries. Most important special cases of Eq.~(\ref%
{ds2CF}) are the de Sitter (dS) and the anti-de Sitter (AdS) spacetimes,
with $\Omega _{\text{dS}}^{2}=1/(Ht)^{2}$ and $\Omega _{\text{AdS}%
}^{2}=1/(kz)^{2}$, described in the inflationary and Poincare coordinates,
respectively. In particular, the results for AdS bulk can be applied to
massless Dirac fields in higher-dimensional Randall-Sundrum-type braneworld
models with two branes and with compact internal spaces.

\section{Casimir densities in carbon nanotubes}

\label{sec:Nano}

In a significant variety of planar condensed matter systems, the low-energy
sector is very well described by the Dirac-like model. A well-known example
is the important case of graphene. In this section we apply general results
obtained above for electrons in  cylindrical nanotubes of finite length.
Recently, carbon nanotubes have attracted a lot of attention due to the
experimental observation in them of a number of novel electronic properties,
what renders them very important for technological purposes. A single-wall
cylindrical nanotube is a graphene sheet rolled into a cylindrical shape.
The low-energy excitations of the electronic subsystem in a graphene sheet
can be described by a pair of two-component spinors, $\psi _{A}$ and $\psi
_{B}$, corresponding to the two different triangular sublattices of the
honeycomb lattice of graphene (see, for instance, \cite{Sait98,Seme84}). The
Dirac equation for these spinors has the form%
\begin{equation}
(iv_{F}^{-1}\gamma ^{0}D_{0}+i\gamma ^{l}D_{l}-m)\psi _{J}=0,\
\label{DeqGraph}
\end{equation}%
where $J=A,B$, $l=1,2$, and $D_{\mu }=\partial _{\mu }+ieA_{\mu }$ with $%
e=-|e|$ for electrons. In Eq.~(\ref{DeqGraph}), $v_{F}\approx 10^{8}$ cm/s
represents the Fermi velocity which plays the role of the speed of light. To
make the treatment more general, we have included in Eq.~(\ref{DeqGraph})
the mass (gap) term. The gap in the energy spectrum is essential for many
physical applications. It can be generated by a number of mechanisms (see,
for example, \cite{Seme84,Gusy95,Cham00,Giov07,Seme08}). In particular, they
include the breaking of symmetry between two sublattices by introducing a
staggered onsite energy \cite{Seme84}, the phenomenon of magnetic catalysis
\cite{Gusy95}, and the deformations of bonds in the graphene lattice \cite%
{Cham00}. Another approach is to attach a graphene monolayer to a
substrate, the interaction with which breaks the sublattice
symmetry \cite{Giov07}. Note that the Casimir interaction between
graphene sheets, resulting from the quantum fluctuations of the
bulk electromagnetic field, has been recently investigated in \
Ref.~\cite{Bord06} (for the comparison of the results based on the
hydrodynamic and Dirac models of dispersion for graphene, see
Ref.~\cite{Chur10}).

For the geometry of a carbon nanotube the spatial topology is $R^{1}\times
S^{1}$. The nanotube is characterized by its chiral vector $\mathbf{C}%
_{h}=n_{w}\mathbf{a}_{1}+m_{w}\mathbf{a}_{2}$, where $n_{w}$, $m_{w}$ are
integers, $\mathbf{a}_{1}$ and $\mathbf{a}_{2}$ are the basis vectors of the
hexagonal lattice of graphene and $a=|\mathbf{a}_{1}|=|\mathbf{a}_{2}|=2.46%
\mathring{A}$ is the lattice constant. For the length of the compact
dimension, one has $L=|\mathbf{C}_{h}|=a\sqrt{n_{w}^{2}+m_{w}^{2}+n_{w}m_{w}}
$, where for zigzag and armchair nanotubes, $\mathbf{C}_{h}=(n_{w},0)$ and $%
\mathbf{C}_{h}=(n_{w},n_{w})$, respectively. All other cases correspond to
chiral nanotubes. In the case $n_{w}-m_{w}=3q_{w}$, $q_{w}\in Z$, the
nanotube will be metallic and in the case $n_{w}-m_{w}\neq 3q_{w}$ the
nanotube will be a semiconductor with an energy gap inversely proportional
to its diameter. We will assume that the nanotube has finite length, $a$. As
the Dirac field lives on the cylinder surface, it is natural to impose bag
boundary conditions (\ref{BagCond}) on the cylinder edges which insure a
zero fermion flux through these edges. The additional confinement of the
fermionic field along the tube axis leads to the change of the VEVs. The
corresponding expressions for the fermionic condensate and the
energy-momentum tensor are obtained from the formulas of the previous
sections, taking $D=2$, $p=0$, $q=1$. The periodicity condition along the
compact dimension for the fields $\psi _{J}$ depends on the chirality of the
nanotube. For metallic nanotubes, we have periodic boundary condition ($%
\alpha _{l}=0$) and for semiconductor nanotubes, depending on the chiral
vector, there are two classes of inequivalent boundary conditions,
corresponding to $\alpha _{l}=\pm 1/3$. These phases have opposite signs for
the sublattices $A$ and $B$. The presence of the gauge field in Eq.~(\ref%
{DeqGraph}) leads to an Aharonov-Bohm effect in carbon nanotubes \cite%
{AhBohm}. This effect manifests itself in a periodic energy gap modulation
and conductance oscillations, as a function of the enclosed magnetic flux,
with a period of the order of the flux quantum. As we will see below,
similar oscillations arise for the fermionic condensate and Casimir
densities.

First, we consider the fermionic condensate (see
Ref.~\cite{Beze10} for the fermionic condensate and VEV of the
fermionic current in a $(2+1)$-dimensional conical spacetime, in
the presence of a circular boundary). The corresponding expression
is obtained from the formulas in Sect.~\ref{sec:FC} taking $p=0$,
$D=2$ and summing the contributions coming from the two
sublattices with opposite signs of $\alpha _{2}\equiv \alpha $.
For an infinite carbon nanotube, from Eq.~(\ref{FC0b}) one has,
for the pure topological part,
\begin{equation}
\langle \bar{\psi}\psi \rangle _{\text{cn}}^{(0)}=-\frac{2mv_{F}}{\pi L}%
\sum_{n=1}^{\infty }\frac{e^{-nmL}}{n}\cos (2\pi n\alpha )\cos (2\pi n\Phi
/\Phi _{0}),  \label{FC0cn}
\end{equation}%
where $\Phi =A_{2}L$ is the magnetic flux across the nanotube and $\alpha
=0,1/3$ for a metallic and for a semiconducting nanotube, respectively. In
the case of a nanotube of finite length, $a$, the general expression (\ref%
{DeltaFC1}) takes the form%
\begin{equation}
\langle \bar{\psi}\psi \rangle _{\text{cn}}=\langle \bar{\psi}\psi \rangle _{%
\text{cn}}^{(0)}-\frac{v_{F}}{\pi L}\sum_{n=-\infty }^{+\infty
}\sum_{j=+,-}\int_{m_{n}^{(j)}}^{\infty }dx\frac{%
(m+x)(e^{2xz}+e^{2ax-2xz})-2m}{\sqrt{x^{2}-m_{n}^{(j)2}}\left( \frac{x+m}{x-m%
}e^{2ax}+1\right) },  \label{FCcn}
\end{equation}%
with the notation%
\begin{equation}
m_{n}^{(\pm )}=\sqrt{k_{n}^{(\pm )2}+m^{2}},\;k_{n}^{(\pm )}=2\pi |n+\alpha
\pm \Phi /\Phi _{0}|/L.  \label{mn}
\end{equation}%
Note that for metallic nanotubes the contribution of the terms with $j=+$
and $j=-$ coincide. In the absence of the magnetic flux, the sublattices
give the same contribution to the condensate. For a massless field the
purely topological part vanishes and the boundary induced part reduces to%
\begin{equation}
\langle \bar{\psi}\psi \rangle _{\text{cn}}=\frac{1}{\pi L}\sum_{n=-\infty
}^{+\infty }\sum_{j=+,-}k_{n}^{(j)}\sum_{l=1}^{\infty
}(-1)^{l}[K_{1}(2k_{n}^{(j)}(al-z))+K_{1}(2k_{n}^{(j)}(al-a+z))].
\label{FCcnm0}
\end{equation}%
As already mentioned, the condensate is a periodic function of the magnetic
flux, with period equal to the flux quantum $\Phi _{0}$. Various important
limiting cases directly follow from the analysis given above.

Now we turn to the VEV of the energy-momentum tensor. For the pure
topological part, we have the expression (no summation over $\mu $)%
\begin{equation}
\langle T_{\mu }^{\nu }\rangle _{\text{cn}}^{(0)}=\frac{2v_{F}\delta _{\mu
}^{\nu }}{\pi L^{3}}\sum_{n=1}^{\infty }\cos (2\pi n\alpha )\cos (2\pi n\Phi
/\Phi _{0})C_{\mu }(nmL)\frac{e^{-nmL}}{n^{3}},  \label{EMT0cn}
\end{equation}%
with the notations%
\begin{equation}
C_{0}(x)=C_{1}(x)=1+x,\;C_{2}(x)=-2-2x-x^{2}.  \label{Clcn}
\end{equation}%
In particular, in absence of magnetic flux, the corresponding energy density
is positive for metallic nanotubes and negative for semiconducting ones.
This means that, from the topological viewpoint, semiconducting nanotubes
are more stable. For a finite length nanotube, from Eq.~(\ref{EMT2pl}) we
find the following expression (no summation over $\mu $)%
\begin{equation}
\langle T_{\mu }^{\nu }\rangle _{\text{cn}}=\langle T_{\mu }^{\nu }\rangle _{%
\text{cn}}^{(0)}-\frac{v_{F}\delta _{\mu }^{\nu }}{\pi L}\sum_{n=-\infty
}^{+\infty }\sum_{j=+,-}\int_{m_{n}^{(j)}}^{\infty }dx\,\frac{%
(x^{2}-m_{n}^{(j)2})^{-1/2}}{\frac{x+m}{x-m}e^{2ax}+1}G_{\mu }^{(j)}(x),
\label{EMTcn}
\end{equation}%
where%
\begin{eqnarray}
G_{0}^{(j)}(x) &=&(x^{2}-m_{n}^{(j)2})\left[ 2+\frac{m}{x-m}%
(e^{2xz}+e^{2ax-2xz})\right] ,  \notag \\
G_{1}^{(j)}(x) &=&-2x^{2},  \label{Glcn} \\
G_{2}^{(j)}(x) &=&k_{n}^{(j)2}\left[ 2+\frac{m}{x-m}(e^{2xz}+e^{2ax-2xz})%
\right] .  \notag
\end{eqnarray}%
In absence of magnetic flux, the energy density corresponding to Eq.~(\ref%
{EMTcn}) is always negative for semiconducting nanotubes. For metallic
nanotubes the energy density is positive for long tubes and negative for
short ones. The forces acting on the tube edges are determined by the
component $\langle T_{1}^{1}\rangle _{\text{cn}}$. As already explained
before, the force corresponding to the second term on the rhs of Eq.~(\ref%
{EMTcn}) is attractive, with independence of the nanotube chirality and of
the magnetic flux. This term dominates for short nanotubes. As regards the
first term, it dominates for long tubes and, in the absence of magnetic
flux, it corresponds to the attractive force for metallic nanotubes and to
the repulsive force for semiconducting ones. Hence, in absence of a magnetic
flux, the resulting force is always attractive for metallic nanotubes,
whereas for semiconducting ones the force is attractive for short tubes and
repulsive for long ones. The sign of the force for long nanotubes can be
simply controlled by tuning up the magnetic flux. This result is again of
high technological importance.

\begin{figure}[tbph]
\begin{center}
\begin{tabular}{cc}
\epsfig{figure=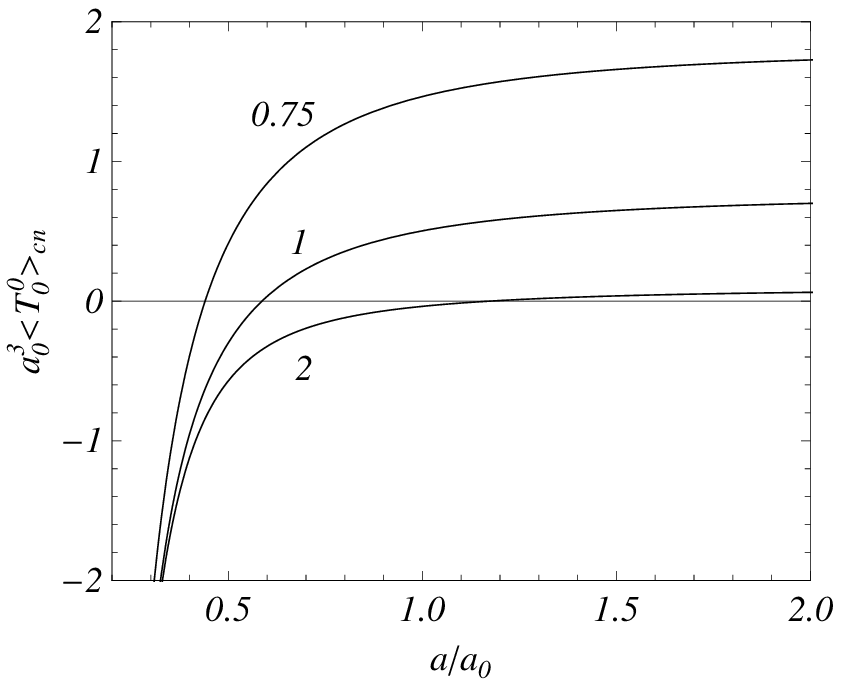,width=7.cm,height=6.cm} & \quad %
\epsfig{figure=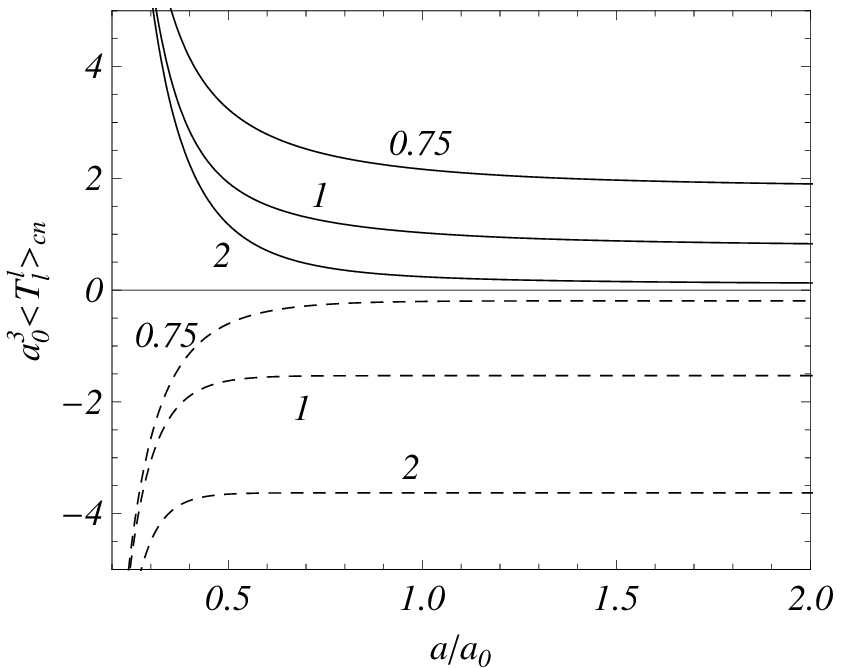,width=7.cm,height=6.cm}%
\end{tabular}%
\end{center}
\caption{VEV of the energy density (left plot) and vacuum stresses (right
plot) for a metallic nanotube, as functions of the tube length, for several
values of its circumference, $L/a_{0}$ (numbers near the curves). On the
left plot full/dashed curves correspond to $l=1/l=2$, respectively.}
\label{fig3}
\end{figure}

In the massless case, the above formulas take the form (no summation over $%
\mu $)%
\begin{equation}
\langle T_{\mu }^{\nu }\rangle _{\text{cn}}=\langle T_{\mu }^{\nu }\rangle _{%
\text{cn}}^{(0)}+\frac{2v_{F}\delta _{\mu }^{\nu }}{\pi L}\sum_{s=1}^{\infty
}\frac{(-1)^{s}}{(2as)^{2}}\sum_{j=+,-}\sum_{n=-\infty }^{+\infty }G_{\mu
}^{(0)}(2sak_{n}^{(j)}),  \label{EMTcnm0}
\end{equation}%
with the notations%
\begin{eqnarray}
G_{0}^{(0)}(x) &=&xK_{1}(x),\;G_{2}^{(0)}(x)=x^{2}K_{0}(x),  \notag \\
G_{1}^{(0)}(x) &=&-xK_{1}(x)-x^{2}K_{0}(x).  \label{Gl0cn}
\end{eqnarray}%
In Fig.~\ref{fig3} we have plotted the vacuum energy density and vacuum
stresses for a massless fermionic field in metallic nanotubes as functions
of the tube length in absence of magnetic flux, $a_{0}$ being an arbitrary
length scale.

For long tubes the pure topological part dominates in the VEV of the energy
density and in the case of the values for parameters corresponding to Fig.~%
\ref{fig3} one has:%
\begin{equation}
\langle T_{l}^{l}\rangle _{\text{cn}}\approx \langle T_{l}^{l}\rangle _{%
\text{cn}}^{(0)}=\frac{2v_{F}\zeta (3)}{\pi L^{3}}\text{diag}(1,1,-2),
\label{TLLcnMlim}
\end{equation}%
with $\zeta (3)\approx 1.202$. The effective pressure on the tube edges is
defined by $-\langle T_{1}^{1}\rangle _{\text{cn}}$ and, as seen from the
right plot, it is always negative, corresponding to attractive forces
between the edges for metallic nanotubes.

\begin{figure}[tbph]
\begin{center}
\begin{tabular}{cc}
\epsfig{figure=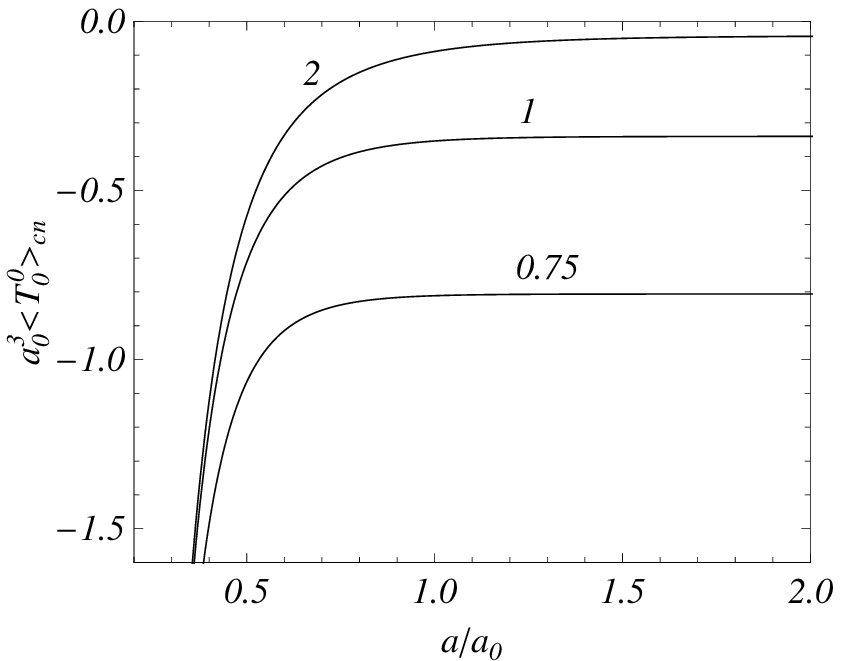,width=7.cm,height=6.cm} & \quad %
\epsfig{figure=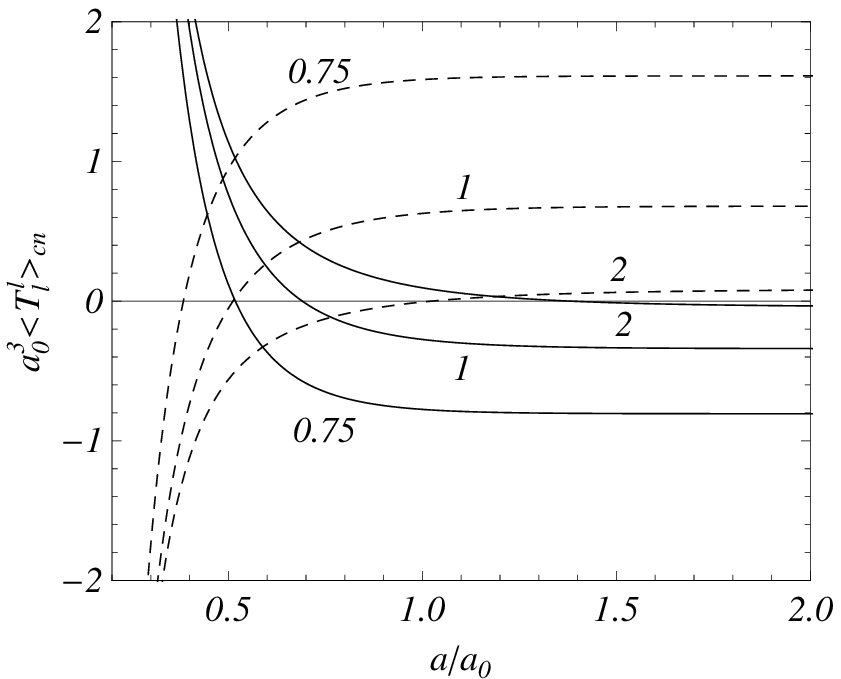,width=7.cm,height=6.cm}%
\end{tabular}%
\end{center}
\caption{The same as in figure \protect\ref{fig3} for semiconducting
nanotube ($\protect\alpha =1/3$).}
\label{fig4}
\end{figure}

Fig.~\ref{fig4} is a corresponding plot of the VEVs of the energy density
and stresses for semiconducting nanotubes ($\alpha =1/3$), for a massless
field in the absence of magnetic flux. Now, for long tubes we have the
limiting value%
\begin{equation}
\langle T_{l}^{l}\rangle _{\text{cn}}\approx \langle T_{l}^{l}\rangle _{%
\text{cn}}^{(0)}=-\frac{0.340v_{F}}{L^{3}}\text{diag}(1,1,-2).
\label{TLLcnSlim}
\end{equation}%
The forces acting on the tube edges are attractive at small distances and
repulsive at large distances.

\begin{figure}[tbph]
\begin{center}
\begin{tabular}{cc}
\epsfig{figure=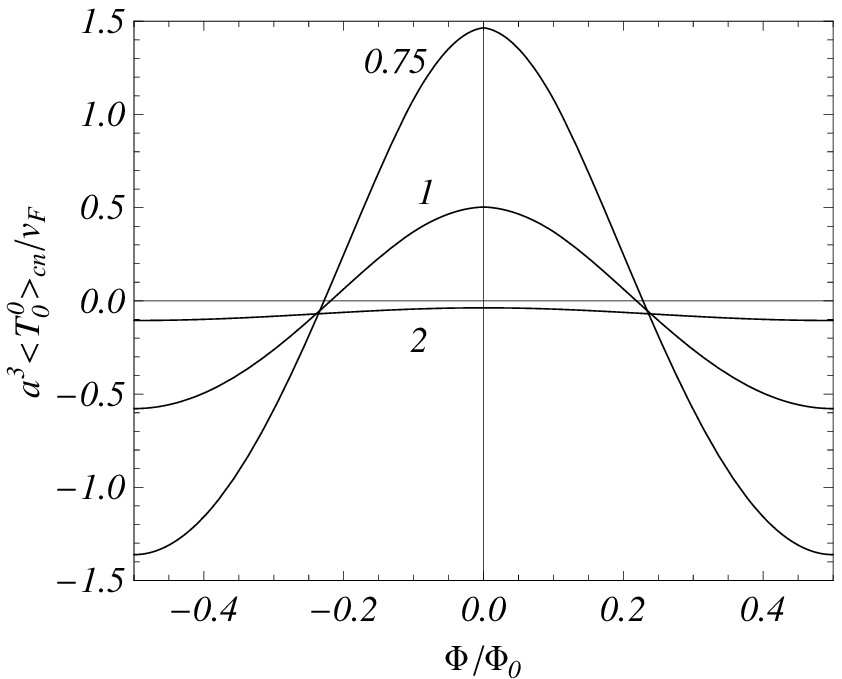,width=7.cm,height=6.cm} & \quad %
\epsfig{figure=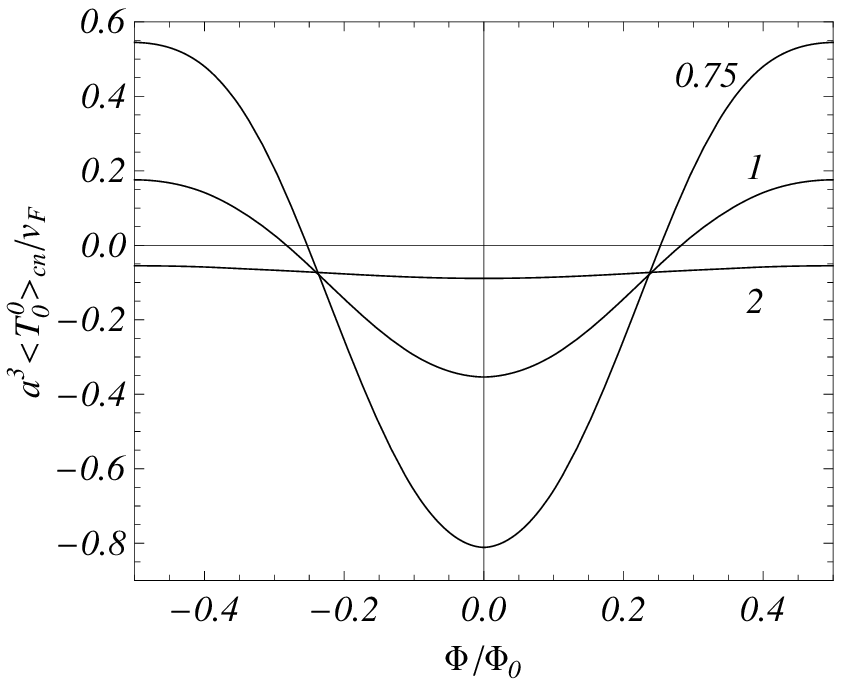,width=7.cm,height=6.cm}%
\end{tabular}%
\end{center}
\caption{VEV of the energy density for metallic (left plot) and
semiconducting (right plot) nanotubes, as functions of the magnetic flux in
units of the flux quantum. The curves correspond to different values of the
ratio $L/a$ (labels of the curves).}
\label{fig5}
\end{figure}

In Fig.~\ref{fig5} we have plotted the vacuum energy density as a function
of the magnetic flux for metallic (left plot) and semiconducting (right
plot) nanotubes, respectively. Recall that the vacuum densities are periodic
functions of the magnetic flux, with period equal to the flux quantum. The
numbers labelling the curves correspond to the values of the ratio $L/a$.

\section{Conclusions}

\label{sec:Conc}

In this paper we have considered the effect of compact spatial dimensions on
the fermionic condensate and VEV of the energy-momentum tensor for a massive
fermion field in the geometry of two parallel plates on which the field
obeys MIT bag boundary condition. Along the compact dimensions, we have
assumed periodicity conditions (\ref{BC}) with constant phases $\alpha _{l}$%
. The eigenvalues of the wave-vector component normal to the plates are
roots of the transcendental equation (\ref{kpvalues}). The mode sums for the
fermionic condensate and the energy-momentum tensor contain series over
these eigenvalues. By applying analytic continuation techniques, as the
Abel-Plana-type summation formula, to these series, we have been able to
explicitly extract and separate, in a cutoff independent way, the purely
topological part and the contributions induced by the single plates. Purely
topological contributions were investigated in Ref.~\cite{Bell09b}, and here
we have been mainly concerned with the boundary induced parts.

In a $(D+1)$-dimensional spacetime with spatial topology $R^{p+1}\times
(S^{1})^{q}$ and with two parallel boundaries, the fermionic condensate is
given by Eq.~(\ref{DeltaFC1}), there the first term on the rhs is a purely
topological contribution and the second term is induced by the presence of
the plates. The boundary induced part is always negative, whereas the purely
topological one, given by Eq.~(\ref{FC0b}), can be either positive or
negative, depending on the values of the phases in the periodicity
conditions along the compact dimensions. For a massless field, the general
formula for the fermionic condensate could be further simplified to Eq.~(\ref%
{FCm0}). Extracting there the contributions corresponding to the geometry of
a single plate, the fermionic condensate could also be presented in the
alternative form (\ref{FCInterf}). The fermionic condensate induced by a
single plate, located at $z=0$, is given by Eq.~(\ref{FC(1)2}). The
condensate diverges on the boundary. For points near the boundary, the
leading term in the asymptotic expansion over the distance from the plate is
given by Eq.~(\ref{FC1Near}). This term does not depend on the lengths of
the compact dimensions and coincides with the boundary induced contribution
of the fermionic condensate for a single plate, in a space with trivial
topology $R^{D}$. Far from the plate, the asymptotic behavior of the
fermionic condensate essentially depends on the phases present in the
periodicity conditions along the compact dimensions.

For $\alpha _{l}=0$, $l=p+2,\ldots ,D$, the main contribution comes from the
zero mode and the quantity $V_{q}\langle \bar{\psi}\psi \rangle
_{p,q}^{(1)}/N_{D}$ coincides with the corresponding result for a plate in
topologically trivial $(p+1)$-dimensional space, $R^{p+1}$. For $\alpha
_{l}\neq 0$, the zero mode is absent and the boundary induced part in the
fermionic condensate is suppressed by the factor $e^{-2m_{0}z}$, where $m_{0}
$ is defined in Eq.~(\ref{m0}). The interference part in the representation (%
\ref{FCInterf}) for the fermionic condensate is given by Eqs.~(\ref%
{FCInterf2}) and (\ref{DeltaFC1m0}) for massive and for massless fields,
respectively. Surface divergences are contained in the single plate parts
and the interference contribution is finite and positive everywhere.

The VEV of the energy-momentum tensor was investigated in Sec. \ref{sec:EMT}%
. In the region between the plates, this VEV is given by Eq.~(\ref{EMT2pl}),
where the second term on the rhs is induced by the plates. The vacuum
stresses along the uncompactified dimensions parallel to the plates are
equal to the energy density. The boundary induced contribution to the energy
density and stresses along the compact dimensions are always negative,
whereas the corresponding contribution to the stress normal to the plates is
positive and, moreover, a uniform function. For a massive field, the energy
density and stresses along the directions parallel to the plates depend on
the distance from the plate and diverge on the boundaries. For a massless
field the VEV of the energy-momentum tensor is uniform, and given by Eq.~(%
\ref{DeltaEMTm0}). An alternative representation for the VEV of the
energy-momentum tensor is (\ref{EMTDec}), with the interference part defined
in Eq.~(\ref{DeltEMT}). The single plate contributions, in this
decomposition, are given by (\ref{EMT1pl1}). The surface divergences in the
VEVs are contained in the single plate parts and the interference part is
finite everywhere. Near the plates, the VEV of the energy-momentum tensor is
dominated by the single plate parts and the leading term in the
corresponding asymptotic expansion is given by Eq.~(\ref{EMT1plNear}). In
the geometry of a single plate and at large distances from the plate,
compared to the length of the compact dimension, the behavior of the VEV is
essentially different and crucially depends on the phases $\alpha _{l}$
included in the periodicity conditions. Thus, for periodic boundary
conditions, $\alpha _{l}=0$, the dominant contribution comes from the zero
mode and for the components along the uncompactified dimensions one has (no
summation over $\mu $) $\langle T_{\mu }^{\mu }\rangle ^{(1)}\approx
N_{D}\langle T_{\mu }^{\mu }\rangle _{R^{p+1}}^{(1)}/(V_{q}N_{p+1})$, $\mu
=0,\ldots ,p+1$, where $\langle T_{\mu }^{\nu }\rangle _{R^{p+1}}^{(1)}$ is
the corresponding VEV for plates in the space $R^{p+1}$. The stress along
the $l$-th compact dimension is suppressed by the factor $e^{-2z/L_{l}}$.
For $\alpha _{l}\neq 0$ all components of the vacuum energy-momentum tensor
are suppressed by the factor $e^{-2m_{0}z}$.

The formulas derived for the fermionic condensate and for the VEV of the
energy-momentum tensor are easily generalized to the case when a constant
gauge field is present. In spite of the fact that the corresponding magnetic
field strength vanishes, the non-trivial topology of the configuration leads
to an Aharonov-Bohm-type effect on the vacuum expectation values. The
corresponding formulas are obtained from those in the absence of a gauge
field by the replacement (\ref{PhaseRepl}).

The vacuum force per unit surface of the plates is determined by
the normal stress and, for the region between the plates, it is
given by Eq.~(\ref{P}), where the first term is the pure
topological part and the second one the interaction part of the
force. When the quantum field lives in all regions of space, the
pure topological parts of the force acting from the left- and from
the right-hand sides of every plate compensate each other, and the
resulting force is just determined by the interaction part. This
turns out to be always attractive, with independence of the
periodicity conditions along the compact dimensions and the value
of the gauge potential. In more physical situations, however, when
the quantum field is confined in the region between the plates,
the pure topological part contributes to the resulting force as
well and in these cases the final forces can be either attractive
or repulsive. Remarkably, the Casimir effect can then be used as a
stabilization mechanism for both the interplate distance and for
the size of the compact subspace in Kaluza-Klein-type models and
in braneworld theories.

For massless fermionic fields, the Casimir densities for the two-plate
geometry in conformally-flat spacetimes with compact spatial dimensions are
obtained from the results above simply by using standard conformal
transformation techniques. The corresponding fermionic condensate and the
VEV of the energy-momentum tensor are thus given by Eq.~(\ref{FCTilCF}),
where the separate contributions: the one induced by non-trivial topology
and the other by the boundary conditions, are given by the second terms on
the right-hand sides. The most important examples of configurations of this
type are dS and AdS spacetimes.

In Sect.~\ref{sec:Nano}, we have applied our general results to the
electrons in finite-length graphene nanotubes. The long-wavelength
excitations of this system is basically described by Dirac's theory, with
the Fermi velocity playing the role of speed of light. The corresponding
expressions for the fermionic condensate and for the Casimir densities are
obtained from the results of the preceding sections by summing the
contributions of the two triangular sublattices of the honeycomb lattice of
the graphene sheet, with opposite signs of the phases along the compact
dimension. In the presence of a magnetic flux, the fermion condensate is
given by Eqs.~(\ref{FCcn}) and (\ref{FCcnm0}), for massive and for massless
fields, respectively. The pure topological contribution is then given by (%
\ref{FC0cn}). In these formulas, the values $\alpha =0$ and $\alpha =1/3$
correspond to metallic and to semiconducting nanotubes, respectively. The
corresponding expressions for the VEV of the energy-momentum tensor have the
form (\ref{EMTcn}) and (\ref{EMTcnm0}). The VEVs are periodic functions of
the magnetic flux with period equal to the flux quantum. In the absence of
magnetic flux, the Casimir forces acting on the edges of the nanotube are
always attractive for metallic nanotubes. However, for semiconductor
nanotubes the forces are attractive for short tubes and repulsive for longer
ones. In the presence of the magnetic flux, the sign of the force for long
tubes can be controlled by tuning the flux. This possibility opens the way
to the design of efficient actuators driven by the Casimir force at the
nanoscale, what is a most cherished goal of present day technology.

\section*{Acknowledgments}

A.A.S. was supported by the ESF Programme \textquotedblleft New Trends and
Applications of the Casimir Effect". E.E. and S.D.O. were partially funded
by MICINN (Spain) project FIS2006-02842, by CPAN Consolider Ingenio Project,
and by AGAUR (Catalonia) 2009SGR-994.


\begin{thebibliography}{99}
\bibitem{Lind04} A. Linde, JCAP \textbf{0410}, 004 (2004).

\bibitem{Sait98} R. Saito, G. Dresselhaus, and M. S. Dresselhaus, \textit{%
Physical Properties of Carbon Nanotubes} (Imperial College Press, London,
1998); C. Dupas, P. Houdy, and M. Lahmani (Editors), \textit{Nanoscience:
Nanotechnologies and Nanophysics} (Springer, Berlin, 2007).

\bibitem{Seme84} G.W. Semenoff, Phys. Rev. Lett. \textbf{53}, 2449 (1984).

\bibitem{Vinc84} D.P. Di Vincenzo and E.J. Mele, Phys. Rev. B \textbf{29},
1685 (1984); J. Gonz\`{a}lez, F. Guinea, and M.A.H. Vozmediano, Nucl. Phys.
B \textbf{406}, 771 (1993); Phys. Rev. B \textbf{63}, 134421 (2001); H.-W.
Lee and D.S. Novikov, Phys. Rev. B \textbf{68}, 155402 (2003); S.G.
Sharapov, V.P. Gusynin, and H. Beck, Phys. Rev. B \textbf{69}, 075104
(2004); K.S. Novoselov et al, Nature \textbf{438}, 197 (2005); D. S. Novikov
and L. S. Levitov, Phys. Rev. Lett. \textbf{96}, 036402 (2006); E. Perfetto,
J. Gonz\'{a}lez, F. Guinea, S. Bellucci, and P. Onorato, Phys. Rev. B
\textbf{76}, 125430 (2007); V.P. Gusynin, S.G. Sharapov, and J.P. Carbotte,
Int. J. Mod. Phys. B \textbf{21}, 4611 (2007); A.H. Castro Neto, F. Guinea,
N.M.R. Peres, K.S. Novoselov, and A.K. Geim, Rev. Mod. Phys. \textbf{81},
109 (2009).

\bibitem{Most97} E. Elizalde, S.D. Odintsov, A. Romeo, A.A. Bytsenko and S.
Zerbini, \textit{Zeta regularization techniques with applications} (World
Scientific, Singapore, 1994); E. Elizalde, \textit{Ten physical applications
of spectral zeta functions}, Lecture Notes in Physics (Springer-Verlag,
Berlin, 1995); V.M. Mostepanenko and N.N. Trunov, \textit{The Casimir Effect
and Its Applications} (Clarendon, Oxford, 1997); M. Bordag, U. Mohidden, and
V.M. Mostepanenko, Phys. Rep. \textbf{353}, 1 (2001); K.A. Milton, \textit{%
The Casimir Effect: Physical Manifestation of Zero-Point Energy} (World
Scientific, Singapore, 2002); M. Bordag, G.L. Klimchitskaya, U. Mohideen,
and V.M. Mostepanenko, \textit{Advances in the Casimir Effect} (Oxford
University Press, Oxford, 2009); M.J. Duff, B.E.W. Nilsson, and C.N. Pope,
Phys. Rep. \textbf{130}, 1 (1986); A.A. Bytsenko, G. Cognola, L. Vanzo, and
S. Zerbini, Phys. Rep. \textbf{266}, 1 (1996).

\bibitem{Buch89} I.L. Buchbinder and S.D. Odintsov, Fortsch. Phys. \textbf{37%
}, 225 (1989).

\bibitem{Eliz01} E. Elizalde, Phys. Lett. B \textbf{516}, 143 (2001); C.L.
Gardner, Phys. Lett. B \textbf{524}, 21 (2002); K.A. Milton, Grav. Cosmol.
\textbf{9}, 66 (2003); E. Elizalde, J. Phys. A \textbf{39}, 6299 (2006); B.
Green and J. Levin, J. High Energy Phys. \textbf{11}, 096 (2007); P.
Burikham, A. Chatrabhuti, P. Patcharamaneepakorn, and K. Pimsamarn, J. High
Energy Phys. \textbf{07}, 013 (2008).

\bibitem{Most87} V.M. Mostepanenko and I.Yu. Sokolov, Phys. Lett. A \textbf{%
125}, 405 (1987); J. C. Long, H. W. Chan and J. C. Price, Nucl. Phys. B
\textbf{539}, 23 (1999); R.S. Decca, D. L\'{o}pez, E. Fischbach, G.L.
Klimchitskaya, D.E. Krause, and V.M. Mostepanenko, Phys. Rev. D \textbf{75},
077101 (2007); G.L. Klimchitskaya, U. Mohidden, and V.M. Mostepanenko, Rev.
Mod. Phys. \textbf{81}, 1827 (2009).

\bibitem{Chen06} H.B. Cheng, Phys. Lett. B \textbf{643}, 311 (2006); H.B.
Cheng, Phys. Lett. B \textbf{668}, 72 (2008); S.A. Fulling and K. Kirsten,
Phys. Lett. B \textbf{671}, 179 (2009); K. Kirsten and S.A. Fulling, Phys.
Rev. D \textbf{79}, 065019 (2009); E. Elizalde, S.D. Odintsov, and A.A.
Saharian, Phys. Rev. D \textbf{79}, 065023 (2009); L.P. Teo, Phys. Lett. B
\textbf{672}, 190 (2009); L.P. Teo, Nucl. Phys. B \textbf{819}, 431 (2009);
L.P. Teo, JHEP \textbf{0906}, 076 (2009); L.P. Teo, JHEP \textbf{0911}, 095
(2009).

\bibitem{Popp04} K. Poppenhaeger, S. Hossenfelder, S. Hofmann, and M.
Bleicher, Phys. Lett. B \textbf{582}, 1 (2004); A. Edery and V.N.
Marachevsky, Phys. Rev. D \textbf{78}, 025021 (2008); A. Edery and V.N.
Marachevsky, JHEP \textbf{0812}, 035 (2008); F. Pascoal, L.F.A. Oliveira,
F.S.S. Rosa, and C. Farina, Braz. J. Phys. \textbf{38}, 581 (2008); L.
Perivolaropoulos, Phys. Rev. D \textbf{77}, 107301 (2008).

\bibitem{Brane} V.A. Rubakov, Phys. Usp. \textbf{44}, 871 (2001); P. Brax
and C. Van de Bruck, Classical Quantum Gravity \textbf{20}, R201 (2003); R.
Maartens, Living Rev. Relativity \textbf{7}, 7 (2004).

\bibitem{Rand99} L. Randall and R. Sundrum, Phys. Rev. Lett. \textbf{83},
3370 (1999).

\bibitem{Gold00} W. Goldberger and I. Rothstein, Phys. Lett. B \textbf{491},
339 (2000); S. Nojiri, S.D. Odintsov, and S. Zerbini, Classical Quantum
Gravity \textbf{17}, 4855 (2000); A. Flachi and D. J. Toms, Nucl. Phys. B
\textbf{610}, 144 (2001); J. Garriga, O. Pujol\`{a}s, and T. Tanaka, Nucl.
Phys. B \textbf{605}, 192 (2001); E. Elizalde, S. Nojiri, S.D. Odintsov, and
S. Ogushi, Phys. Rev. D \textbf{67}, 063515 (2003); A.A. Saharian and M.R.
Setare, Phys. Lett. B \textbf{584}, 306 (2004); A. Flachi, A. Knapman, W.
Naylor, and M. Sasaki, Phys. Rev. D \textbf{70}, 124011 (2004); M. Frank, I.
Turan, and L. Ziegler, Phys. Rev. D \textbf{76}, 015008 (2007); L.P. Teo,
Phys. Lett. B \textbf{682}, 259 (2009); A. Flachi and T. Tanaka, Phys. Rev.
D \textbf{80}, 124022 (2009); L.P. Teo, Phys. Rev. D \textbf{82}, 027902
(2010); L.P. Teo, JHEP \textbf{1010}, 019 (2010).

\bibitem{Knap04} A. Knapman and D. J. Toms, Phys. Rev. D \textbf{69}, 044023
(2004); A. A. Saharian, Nucl. Phys. B \textbf{712}, 196 (2005).

\bibitem{Flac03} A. Flachi, J. Garriga, O. Pujol\`{a}s, and T. Tanaka, J.
High Energy Phys. \textbf{0308}, 053 (2003); A. Flachi and O. Pujol\`{a}s,
Phys. Rev. D \textbf{68}, 025023 (2003); A.A. Saharian, Phys. Rev. D \textbf{%
73}, 044012 (2006); A.A. Saharian, Phys. Rev. D \textbf{73}, 064019 (2006);
A.A. Saharian, Phys. Rev. D \textbf{74}, 124009 (2006); R. Linares, H.A.
Morales-T\'{e}cotl, and O. Pedraza, Phys. Rev. D \textbf{77}, 066012 (2008);
M. Frank, N. Saad, and I. Turan, Phys. Rev. D \textbf{78}, 055014 (2008).

\bibitem{Bell09} S. Bellucci and A.A. Saharian, Phys. Rev. D \textbf{80},
105003 (2009).

\bibitem{Inag97} T. Inagaki, T. Muta, and S.D. Odintsov, Prog. Theor. Phys.
Supll. \textbf{127}, 93 (1997).

\bibitem{Flac10} A. Flachi and T. Tanaka, arXiv:1012.0463.

\bibitem{John75} K. Johnson, Acta Phys. Polonica B \textbf{6}, 865 (1975).

\bibitem{Mama80} S. G. Mamaev and N. N. Trunov, Sov. Phys. \textbf{23}, 551
(1980).

\bibitem{Paol99} R. D. M. De Paola, R. B. Rodrigues, and N. F. Svaiter, Mod.
Phys. Lett. A \textbf{14}, 2353 (1999).

\bibitem{Eliz02} E. Elizalde, F.C. Santos, and A.C. Tort, Int. J. Mod. Phys.
A \textbf{18}, 1761 (2003).

\bibitem{Lutk84} C.A. L\"{u}tken and F. Ravndal, J. Phys. G \textbf{10}, 123
(1984); S.A. Gundersen and F. Ravndal, Ann. Phys. \textbf{182}, 90 (1988).

\bibitem{Sund04} P. Sundberg and R.L. Jaffe, Ann. Phys. \textbf{309}, 442
(2004).

\bibitem{Fosc08} C.D. Fosco and E.L. Losada, Phys. Rev. D \textbf{78},
025017 (2008).

\bibitem{Bell09b} S. Bellucci and A.A. Saharian, Phys. Rev. D \textbf{79},
085019 (2009).

\bibitem{Bell10} S. Bellucci, A.A. Saharian, and V.M. Bardeghyan, Phys. Rev.
D \textbf{82}, 065011 (2010).

\bibitem{Birr82} N.D. Birrell and P.C.W. Davis, \textit{Quantum Fields in
Curved Space} (Cambridge University Press, Cambridge, England, 1982).

\bibitem{Rome02} A. Romeo and A. A. Saharian, J. Phys. A: Math. Gen. \textbf{%
35}, 1297 (2002).

\bibitem{Saha08Rev} A. A. Saharian, \textit{The Generalized Abel-Plana
Formula with Applications to Bessel Functions and Casimir Effect} (Yerevan
State University Publishing House, Yerevan, 2008); Preprint ICTP/2007/082;
arXiv:0708.1187.

\bibitem{Saha08} A.A. Saharian, Class. Quantum Grav. \textbf{25}, 165012
(2008); E.R. Bezerra de Mello and A.A. Saharian, J. High Energy Phys.
\textbf{12} (2008) 081.

\bibitem{Saha04} A.A. Saharian, Phys. Rev. D \textbf{69}, 085005 (2004).

\bibitem{Gusy95} V. P. Gusynin, V. A. Miransky, and I. A. Shovkovy, Phys.
Rev. D \textbf{52}, 4718 (1995).

\bibitem{Cham00} C. Chamon, Phys. Rev. B \textbf{62}, 2806 (2000); C.-Y.
Hou, C. Chamon, and C. Mudry, Phys. Rev. Lett. \textbf{98}, 186809 (2007).

\bibitem{Giov07} G. Giovannetti, P.A. Khomyakov, G. Brocks, P.J. Kelly, and
J. van den Brink, Phys. Rev. B \textbf{76}, 073103 (2007); S.Y. Zhou et al.,
Nature Mater. \textbf{6}, 770 (2007).

\bibitem{Seme08} G.W. Semenoff, V. Semenoff, and F. Zhou, Phys. Rev. Lett.
\textbf{101}, 087204 (2008).

\bibitem{Bord06} M. Bordag, B. Geyer, G.L. Klimchitskaya, and V.M.
Mostepanenko, Phys. Rev. B \textbf{74}, 205431 (2006); M. Bordag, I.V.
Fialkovsky, D.M. Gitman, and D.V. Vassilevich, Phys. Rev. B \textbf{80},
245406 (2009); M. Bordag, I.V. Fialkovsky, D.M. Gitman, and D.V.
Vassilevich, arXiv:1003.3380; D. Drosdoff and L.M. Woods, arXiv:1007.1231;
B.E. Sernelius, arXiv:1011.2363.

\bibitem{Chur10} Yu.V. Churkin, A.B. Fedortsov, G.L. Klimchitskaya, and V.A.
Yurova, Phys. Rev. B \textbf{82}, 165433 (2010).

\bibitem{AhBohm} H. Ajiki and T. Ando, J. Phys. Soc. Jpn. \textbf{62}, 1255
(1993); Physica B \textbf{201}, 349 (1994); A. Bachtold et al., Nature
\textbf{397}, 673 (1999); S. Zaric et al., Science \textbf{304}, 1129
(2004); U.S. Coskun et al., Science \textbf{304}, 1132 (2004); J. Cao, Q.
Wang, M. Rolandi, and H. Dai, Phys. Rev. Lett. \textbf{93}, 216803 (2004);
B. Lassagne et al., Phys. Rev. Lett. \textbf{98}, 176802 (2007); M.-G. Kang
et al., Phys. Rev. B \textbf{77}, 113408 (2008).

\bibitem{Beze10} E.R. Bezerra de Mello, V.B. Bezerra, A.A. Saharian, and
V.M. Bardeghyan, Phys. Rev. D \textbf{82}, 085033 (2010); S. Bellucci, E.R.
Bezerra de Mello, A.A. Saharian, arXiv:1101.4130.
\end{thebibliography}
\end{document}